\documentclass[twocolumn,preprintnumbers,amsmath,amssymb,aps,pre,reprint,longbibliography]{revtex4-2}
\usepackage{graphicx}
\usepackage{xcolor}
\begin{document}

\title{Pattern Formation and Stick-Slip Dynamics In Binary Particle Assemblies with Rotating Drives}
\author{
C. Reichhardt and C. J. O. Reichhardt 
} 
\affiliation{
Theoretical Division and Center for Nonlinear Studies,
Los Alamos National Laboratory, Los Alamos, New Mexico 87545, USA
}

\date{\today}

\begin{abstract}
We numerically examine a binary system of particles with repulsive interactions, where one species is driven by a rotating drive and the other is subjected either to a constant drive in a fixed direction or to a rotating drive that is out of phase with the first species. As a function of rotation frequency, we find a variety of order-disorder transitions and pattern forming states, including density-modulated stripes, partially jammed states, phase separated fluids, and mixed fluids. When one species has a constant drive and the drive on the other species is rotated at low frequencies, the system switches between different pattern forming phase-separated lanes including density-modulated stripes and partially jammed states, similar to what is observed for oppositely driven colloids. The lanes tend to align with the net direction of rotation, resulting in a series of order-disorder switching transitions. The transport curves show abrupt jumps up or down at the transitions, which also correspond with changes in the topological order. We find similar switching transitions when both species rotate out of phase with each other. For intermediate driving frequencies, the system becomes increasingly fluid-like and the laning behavior is lost. At high frequencies, however, the system can again exhibit patterned flow when the rotation orbits become smaller than the average spacing between particles. The switching is reduced when a finite temperature is included, but even for temperatures at which the uniform equilibrium bulk system is liquid, the partially jammed state can generate local density enhancements that lead to recrystallization. We demonstrate the pattern switching behavior for systems with different screened repulsive interaction potentials.
\end{abstract}

\maketitle

\section{Introduction}

The formation of patterns such as stripes
can arise in equilibrium systems where the
particles have a competing attraction and repulsion
\cite{Reichhardt10, Meng17, Xu21, Hooshanginejad24}
or multiple-step interaction potentials \cite{Seul92, Malescio03}.
Such patterns can also occur in 
nonequilibrium systems when the interactions are
modified by a time-dependent external field \cite{Massana21, Katzmeier22}
or when some form of patterned substrate
is present \cite{McDermott14, Reichhardt17}.
Nonequilibrium stripe-like patterns
have also been observed in binary systems of 
particles that move with different relative velocities or
move in opposite directions to each other in the
presence of an external field \cite{Dzubiella02a, Rex07,Glanz12}.
Similar patterns can
occur when one species couples to the drive while the other does
not.
The applied drive can be in a fixed direction,
or it can be a time-periodic ac \cite{Wysocki09, Vissers11a,Li21}
or circular drive
\cite{Tierno07, vanZuiden16, Han17, Reichhardt19, Reichhardt19b}.
One of the most studied examples of
stripe pattern formation
is a binary assembly of particles that move in opposite directions. At low drives the system exhibits a fluid or mixed phase,
while at high drives it can form a phase-separated, laned state in which the two species organize into oppositely moving lanes \cite{Dzubiella02a, Chakrabarti04, Rex07, Glanz12, Ikeda12, Klymko16, Poncet17}.

The particles can reduce the frequency of collisions
by phase separating,
and the phase-separated states may be fluid-like or
can have crystalline order \cite{Glanz16,Reichhardt18}.
Depending on the form of the particle-particle interaction and the
amplitude of the drive,
a number of additional phases can appear,
such as jammed states, intermittent states, and different
types of fluids \cite{Helbing00,Glanz16,Bain17,Reichhardt18,Yu22,Yu24}.
Studies of a smaller number of driven particles moving through a
background of nondriven particles showed that
the driven particles experience an effective attraction
to each other \cite{Reichhardt06}.
Laning transitions have been studied for particles with Yukawa or
screened-Coulomb interactions, hard disks, Coulomb interactions, and
particles interacting with a combination of attraction and repulsion
\cite{Kogler15,Wachtler16,Reichhardt26}.
Other systems of oppositely driven particles
that have been investigated include disks \cite{Reichhardt18},
colloids with opposite charges \cite{Leunissen05},
dusty plasma particles \cite{Sutterlin09},
particles with different bounciness in gravitational fields \cite{Isele23},
and active matter \cite{Reichhardt18b,Khelfa22}.
Similar lane formation can occur in models of social dynamics,
such as pedestrian flows moving in opposite directions \cite{Feliciani16},
and tilting of the lanes can be induced when chiral effects are added
\cite{Reichhardt19b,Bacik23}.
Lane formation has been observed for
mixtures of magnetic skyrmions with different topological numbers,
in which the skyrmions move at different speeds or have
different Hall angles \cite{Vizarim25},
as well as for mixtures of magnetic vortices and skyrmions \cite{Neto22}.
Similar dynamic behavior
could arise in bilayers of charged particles \cite{Zarenia17,Zhou21},
where one layer is driven and the other 
is not, or in bilayer colloidal systems \cite{Vezirov13}.

In general, most studies of lane-forming systems have involved a
fixed direction of the drives on the two species, where
the lanes align with the direction of the net drive.
Far less is known about what happens if one or
both of the driving directions change as a function of time.
For oppositely driven particles, if the drives on both species
are rotated in the
same direction at the same rate,
the lane states would rotate smoothly with the drive direction.
If, however, the drive on one species is
held in a fixed direction while the driving direction
of the other species is rotated,
it is not known what would happen.
Possible outcomes include formation of a single lane,
switching among different laned states, or the simple formation of
a disordered flow state.
The behavior
should depend strongly on the rate at which the drive direction changes.
In particular, if the
rate is faster than the speed at which stable
lanes can form, the system will not be able to keep up with the changing
drive.
In a recent study \cite{Reichhardt26},
a binary assembly of particles,
termed species A and B, were
driven perpendicular to one another rather than in opposite directions.
When the drives have the same magnitude,
the particles form lanes oriented at
$45^\circ$ to the longitudinal drive.
If the longitudinal drive amplitude $F^B_D$ is
held fixed along $\bf \hat{x}$
while the perpendicular drive amplitude
$F^A_D$ is increased along $\bf \hat{y}$,
then instead of forming 
a single laned state, the system
exhibits a series of different laned states in which
the lanes are tilted with respect to the longitudinal drive at angles near
$ \theta = \arctan(F^{A}_{D}/F^{B}_{D})$.
A tilted lane state remains stable over a range of $F^A_D$ values, but
when $F^A_D$ has increased too far above the value at which the
lanes formed, the lanes break apart
and the system reforms a new lane state with a higher tilt angle.
The transitions into and out of the different lane states are
accompanied by
sharp jumps up and down in the velocity versus drive curves,
as well as changes in the number of topological defects.
In Ref.~\cite{Reichhardt26}
the perpendicular drive was increased very slowly to ensure that
the system had enough time
to form a lane, and the rate dependence of the
change in drive direction was not explored.

In this work we consider a binary system of particles
with either screened Yukawa, Bessel function, or Coulomb repulsive
interactions, in which
the direction of drive $F^B_D$
on species B
is held fixed while the
drive direction of $F^A_D$ on species
A is rotated at a frequency $\omega$.
For small $\omega$ or
slow rotation,
a series of different ordered pattern-forming lane states appear,
along with
phase separated density-modulated stripe structures
that can be regarded as partially jammed states
in which one species partially blocks the flow of the other.
The ordered states tend to align with the
net direction of the rotating drive; however, as this
direction changes, the patterns break up,
the system becomes liquidlike, and a different lane configuration
forms that is oriented along a new angle.
The transitions between the different lane states
are associated with pronounced jumps in the transport curves
and in the amount of topological order in the system.
The laned states
are partially ordered or crystalline,
whereas at the transitions between laned states
the system is more disordered or fluid-like. 
As $\omega$ increases, the system becomes more
disordered and the lanes become less well defined,
and for high frequencies, the system remains fluidlike because
the particles do not have time to form steady state lanes before
the drive direction changes again. 
At the highest rotation frequencies considered here,
ordered lanes reappear when the radius of the particle orbits
becomes smaller than the average interparticle spacing.

We have also considered the case where 
the drives on both particle species are rotated out of phase.
Compared to rotating only one of the drives, for the two rotated drives
we observe an even larger number of
jumps in the transport curves.
At low frequencies, a series of laned states appear
that have pronounced
switching dynamics which are effectively stick-slip behavior.
When a lane forms, it can be regarded as being
stuck flowing in a fixed direction.
As the drive rotates, the perpendicular forces on each lane increase
and compress the lane, but the flow direction remains stuck.
When the compression becomes large enough, the lane breaks apart,
the stress forces on the original lane disappear,
and the flow reorients into
a new laning direction that is strain-free.
With continued rotation of the drive,
the same sticking and compression process repeats on the new lanes.
In stress-strain systems, this is the equivalent of
moving a spring-block slider with constant velocity on the slider.
In our case, the velocity of the moving slider
is the rotating drive, while the sticking object
is the lane itself.
The stuck state is the slowly compressing lane,
and the slip event is the switch to a new laning direction.
As the driving frequency increases, the lane switching
becomes less pronounced and the system acts
more fluidlike, while at high frequencies
the system can form a phase-separated crystal state
in which the orbits are so small that the system behaves like a
collection of point particles.
We find that there is a critical
driving amplitude below which the system remains a crystal,
and we map out a phase diagram of the different
flow states as a function of frequency versus drive amplitude.
We have also considered finite temperatures and
show that the switching dynamics remains robust
against thermal fluctuations.
Even at temperatures for which the undriven system would melt,
certain lane states or partially jammed states
can appear due to strong localized density modulations
that become high enough along a stripe to induce recrystallization.

\section{Simulation}

We consider a two‑dimensional system of size $L \times L$
with $L=48$ and with periodic boundary conditions in the
$x$- and $y$-directions.
There are $N$ particles evenly split between species A and species B that
experience different driving forces but are otherwise identical.
The particle density is $\rho=N/L^2$ and is set to
$\rho=0.292$ for most of this work.
Unless otherwise noted,
the particles interact via a repulsive Yukawa potential,
$ V(R_{ij}) = C e^{-R_{ij}}/R_{ij}$, where $C$ is a prefactor,
${\bf R}_{i(j)}$ is the position of particle $i(j)$, and
$R_{ij}=|{\bf R}_i-{\bf R}_j|$.
The particles are initially placed in a triangular lattice with every other
lattice site occupied by an A or a B particle to give maximal mixing.
The overdamped equation of motion for particle $i$ is
\begin{equation}
\eta \frac{d {\bf R}_i}{dt} = -\sum_{j\neq i}^{N}\nabla V(R_{ij}) + {\bf F}_{D}^i
\end{equation}
where the damping coefficient is set to $\eta = 1.0$.
The size of the molecular dynamics simulation time step is $\delta t = 0.005$.
After the system is initialized,
we employ one of two driving protocols.
In the first, species A is subjected to a rotating drive
$ {\bf F}^{A}_D(t)=A\sin(\omega t)\hat{\bf y}+A\cos(\omega t)\hat{\bf x}$,
while species B experiences a constant dc driving in the $x$-direction,
${\bf F}^{B}_{D}=F_{D}\hat{\bf x}$ 
In the second, both species are driven with rotating fields
that are out of phase,
with ${\bf F}^A_D(t)$ the same as given above, and
$ {\bf F}^{B}(t)= -A\sin(\omega t)\hat{\bf y} - A\cos(\omega t)\hat{\bf x}$.
In either case, we report time in units of the period $\tau=2\pi/\omega$ of the
ac drive.

We measure the nearly instantaneous particle velocities, averaged over
1000 simulation time steps, in the $x$ and $y$ directions,
$V^{A}_{x} = (2/N)\sum^{N}_{i=1}\delta(\sigma_{i}-1)({\bf v}_{i}\cdot\hat{\bf x})$,
$V^{A}_{y} = (2/N)\sum^{N}_{i=1}\delta(\sigma_{i}-1)({\bf v}_{i}\cdot\hat{\bf y})$,
$V^{B}_{x} = (2/N)\sum^{N}_{i=1}\delta(\sigma_{i})({\bf v}_{i}\cdot\hat{\bf x})$,
and
$V^{B}_{y} = (2/N)\sum^{N}_{i=1}\delta(\sigma_{i})({\bf v}_{i}\cdot\hat{\bf y})$,
where ${\bf v}_{i}$ is the instantaneous velocity of
particle $i$ and
$\sigma_{i}=1$(0) indicates that particle $i$ belongs to species A(B).
We also measure the fraction of sixfold-coordinated particles
without regard to species identity,
$P_{6}= (1/N)\sum_{i=1}^{N}\delta(z_{i}-6),$  
where $z_{i}$ is the coordination number of particle
$i$ obtained from a Voronoi tessellation.
At zero temperature and zero drive, the
repulsive particle-particle interactions favor
the formation of a triangular lattice in which $P_6=1.0$.

In addition to the Yukawa interaction,
we consider two alternative pair potentials.
One is a modified Bessel function,
$V(R_{ij}) = F_0K_{0}(R_{ij})$,
where $F_0$ is a constant.
This function
decays exponentially at large distances and
describes vortex interactions in type-II superconductors as well as
certain screened-charge systems \cite{Reichhardt17}. 
The second is a long-range Coulomb potential,
$ V(R_{ij}) = 1/R_{ij}$.
For this potential 
we employ the Lekner summation technique for computational
efficiency \cite{Lekner91,GronbechJensen97a},
as used previously to study driven charge
motion on random substrates \cite{Reichhardt22} or
periodic one-dimensional substrates \cite{Reichhardt25},
as well as oppositely driven particle mixtures \cite{Reichhardt26}.

\section{Switching Dynamics and Pattern Formation at Low Frequencies}

\begin{figure}
\includegraphics[width=\columnwidth]{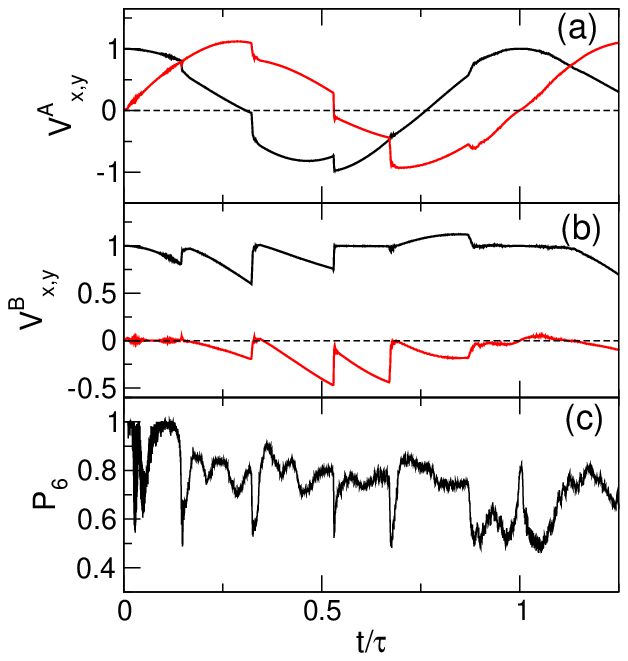}
\caption{Velocity and topological order for a binary assembly of particles.
  The species A rotating drive has
frequency $\omega=2\times 10^{-6}$ and amplitude $A=1.0$,
while species B has a constant drive of $F^{B}_{D} = 1.0$ in the $x$-direction.
(a) Species A velocity $V^{A}_{x}$ (black) and $V^A_y$ (red) vs
time $t/\tau$ where $\tau$ is the drive period. (b) Species B
velocity $V^{B}_{x}$ (black) and $V^B_y$ (red) vs $t/\tau$.
(c) The corresponding $P_{6}$, or species-agnostic fraction of
sixfold-coordinated particles, vs $t/\tau$.
A series of switching events occur, indicated by the jumps in the
velocity of both species, while the velocity
dips correlate with dips in $P_{6}$.}
\label{fig:1}
\end{figure}

We first consider a system of density
$\rho = 0.292$ in which the drive $F^A_D$ on
species A is of amplitude $A=1.0$ and
is rotated in the counterclockwise direction
at a frequency of $\omega = 2\times10^{-6}$,
while species B is subjected to
a constant drive of $F^{B}_{D} = 1.0$ in the $x$-direction.
In Fig.~\ref{fig:1}(a) we plot the $x$ and $y$ velocities 
$V^{A}_x$ and $V^A_y$ versus time $t/\tau$
for
species A, while Fig.~\ref{fig:1}(b) shows the corresponding
$V^B_x$ and $V^B_y$ versus $t/\tau$
for species B. The value of 
the species-independent measure $P_{6}$
is plotted versus $t/\tau$ in Fig.~\ref{fig:1}(c).
For a triangular lattice that rotates as a rigid body,
we would expect smooth sinusoidal 
variation of all four quantities $V^A_x$, $V^A_y$, $V^B_x$, and $V^B_y$
with an additional net translation equal to $F^B_D/2$ in the $x$ direction
for both species, while
$P_6$ would be $1.0$.
Instead, in Fig.~\ref{fig:1},
$V^{A}_{x}$ and $V^A_y$ do not smoothly vary
but contain a series of sudden changes or jumps superimposed on the sinusoidal
variation.
The jump features are more pronounced in $V^B_x$ and $V^B_y$, where they are
superimposed on a background of $V^B_x=1.0$ and $V^B_y=0.0$.
In general, the velocity jumps are correlated with
dips in $P_{6}$ to values close to 0.5,
indicating that the system becomes disordered during
the jumps. Between the jumps, $P_6$ assumes more ordered values
between $0.75$ and $0.9$.

At the low driving frequency used in Fig.~\ref{fig:1},
the system when viewed from the center of mass rest frame is equivalent
to an assembly of oppositely driven particles where the
driving plus
a perpendicular driving component are varied periodically.
If both drives were dc, the particles would
assemble into a stripe or lane-like pattern aligned
in the direction of the net driving force.
Due to the ac drive, this direction keeps changing,
and therefore the lanes must break apart and reform in order
to remain aligned with the drive, resulting in the appearance of a
series of stripe or lane-like patterns.
Once a stripe is formed, it persists over
a range of driving angles, but when the
difference between the drive angle and the lane angle becomes too large,
the system disorders and rearranges into a new lane state.
Within each lane state,
$P_6$ is high and the particle velocities are smoothly changing or are constant.

\begin{figure}
\includegraphics[width=\columnwidth]{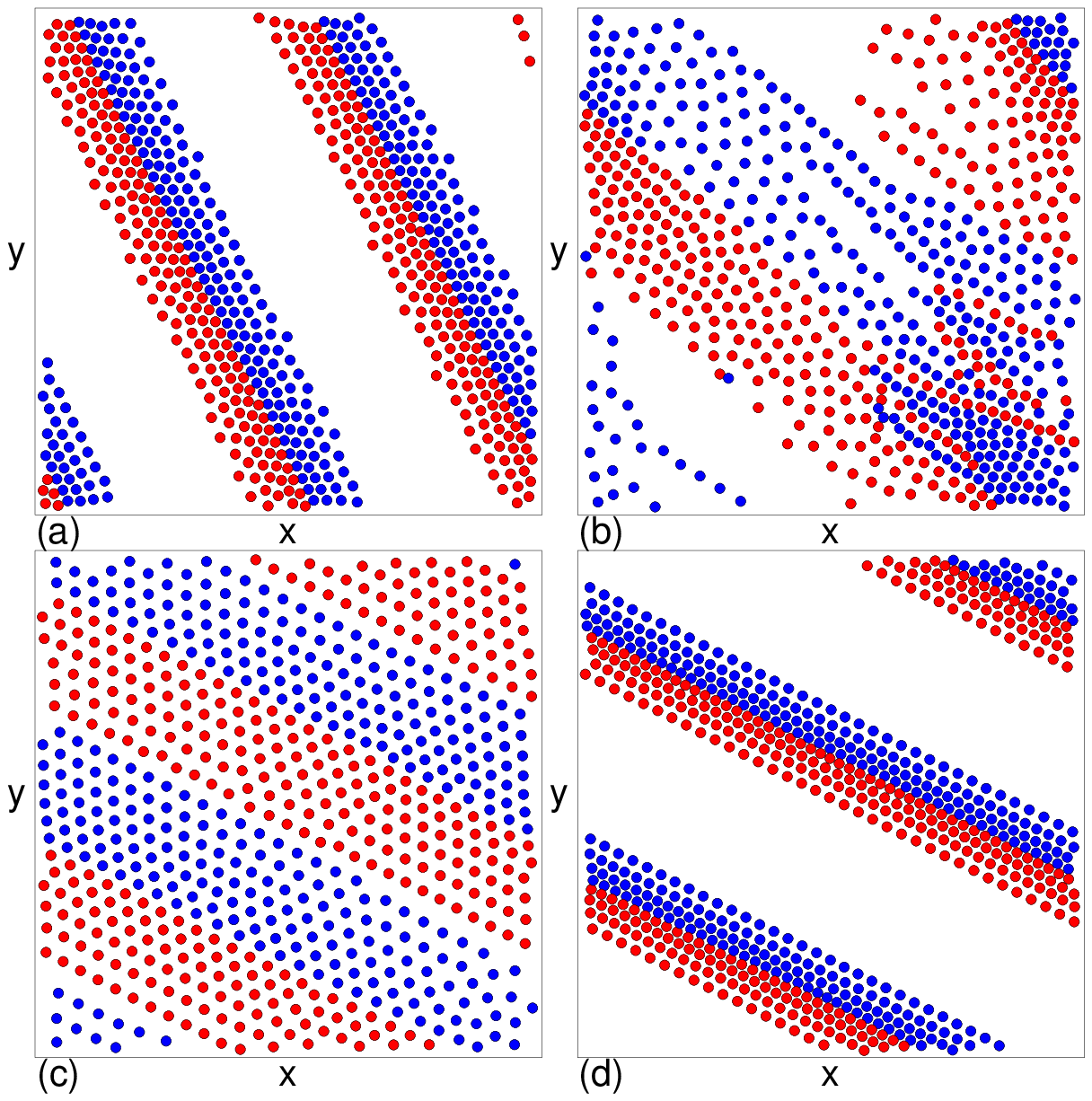}
\caption {Snapshots of particle positions for species A (blue) and B (red)
for the system from Fig.~\ref{fig:1} with
$\omega=2\times 10^{-6}$, $A=1.0$, and constant B species drive $F_D^B=1.0$.
(a) A tilted compressed lane state at $t/\tau=0.31$.
(b) The switching transition at $t/\tau=0.3225$,
where the stripe partially breaks apart.
(c) A new tilted lane state with uniform density at $t/\tau=0.347$.
(d) The same lane state from (c) at $t/\tau=0.5$, where the lanes have
become compressed due to the rotation of the drive.
}
\label{fig:2}
\end{figure}

In Fig.~\ref{fig:2}(a) we show the particle positions for the system
from Fig.~\ref{fig:1} at $t/\tau=0.31$, just before a switching event,
where the system forms a tilted lane or stripe state
oriented close to $120^\circ$ from the $x$-axis.
In this portion of the ac drive cycle, species A experiences
a large drive in $+y$ and a weak drive in $+x$.
As a result, $V^A_x < V^B_x$, so species B is able to
overtake species A along the $x$ direction. As
the stripe of species B particles
pushes up against the tilted stripe of species A particles, it is deflected
along the $-y$ direction, giving a negative
value of $V_B^y$; meanwhile, the species A stripe is moving
primarily in the $+y$ direction.
Once the difference in $+x$ velocities between species A and species
B becomes large enough, the stripes of each species are no longer able to
pass through each other but undergo a clogging event related to 
that observed for oppositely driven particles. Due to the
continually changing species A drive angle, this clog is not
motionless; instead, in the center of mass reference frame of each
stripe, the boundary between species A and species B particles remains
stable while the particles themselves slide perpendicularly to this
boundary in opposite directions from each other.
Since species A is moving more slowly along $+x$ than 
species B, a net compression of the stripe structure results, and
this compression becomes stronger as the $x$ component of the species A drive
decreases with the advancing drive cycle.
The slower motion of species A produces
increased drag on species B, causing
$V^{B}_{x}$ to decrease
with time in this portion of the ac cycle and
giving $V^B_{x} = 0.65$ instead of
the value $V^{B}_{x} = 1.0$ that
would be expected in the absence of species A.
The particles in the inner portion of the stripe maintain a fair amount
of triangular ordering, so that $P_{6}$ is close to $0.8$.
There are still some topological defects present
due to the increased spacing of the particles along the stripe edges.

In Fig.~\ref{fig:2}(b) we show the particle configurations for the
same system from Fig.~\ref{fig:1} at
$t/\tau=0.3225$ where the lane structure breaks apart,
coinciding with a jump up in velocity and
a dip in $P_{6}$.
The overall compression of the system is reduced, and
regions of both low and high density are present.
There is a sudden jump up in species B velocity to
$V^{B}_{x} \approx 1.0$, indicating a strong reduction
in the drag on species B,
while $V^{B}_{y}$ jumps from a finite value to $V^B_y=0.0$.
Jumps of similar magnitude occur in
$V^{A}_{x}$ and $V^{A}_{y}$.
The destruction of the laned state is triggered by the destabilization of
the boundary at the center of the stripe that
separates species A from species B.
This occurs when the
direction of the net velocity difference
between the species deviates too
far from the orientation of the stripe, resulting in a velocity
difference component perpendicular to the A-B boundary that exceeds
a critical value.
Figure~\ref{fig:2}(c)
illustrates the particle configurations at
$t/\tau=0.347$ where a new uniform density tilted lane state has emerged
at a different angle from the previous tilted lane state,
and the system exhibits considerable triangular ordering with $P_6=0.91$.
The new
stripe angle is closer to $180^\circ$ than the previous
stripe angle, permitting
species B with its higher $x$ velocity to move
more easily past species A with
its decreasing $x$ velocity.
This new laned state persists over the range
$t/\tau=0.35$ to $t/\tau=0.52$,
and both $V^{B}_{x}$ and $V^{A}_{y}$ decrease monotonically during this
time interval.

In general, the lane state has a nearly uniform density
just after each switching event, and then becomes increasingly compressed
until the next switching event occurs.
Each time the density becomes uniform, both species of particles revert 
nearly to the velocities they would have had if their motion were aligned with
the net direction of the total force.
The reorientation of the lanes destroys the clogging behavior
until the driving direction
of species A changes again and clogging begins to reemerge.
The
lane compression results from the component of the velocity difference between
species A and species B that is perpendicular to the A-B boundary,
and after the switching event,
this velocity difference is aligned with
the net direction of drive,
leading to the formation of uncompressed
uniform density lanes which then compress
again as the drive angle is rotated further.
In Fig.~\ref{fig:2}(d) we show the particle configurations
for the lane state from Fig.~\ref{fig:2}(c) 
at $t/\tau=0.5$
where the lanes have become strongly compressed.
Here the instantaneous drives are
$F^{A}_{y} = 0.0$ and $F^A_{x} = -1.0$, so that
the ideal laned state would be aligned with the $x$ direction;
however, once the tilted lane state from
Fig.~\ref{fig:2}(c) has formed at $t/\tau=0.347$, the system remains locked
in this orientation since 
there is strong memory of the previous orientation of the net driving
force. Thus instead of an $x$-aligned set of lanes at $t/\tau=0.5$, we
find the highly compressed tilted lanes of Fig.~\ref{fig:2}(d).

\begin{figure}
\includegraphics[width=\columnwidth]{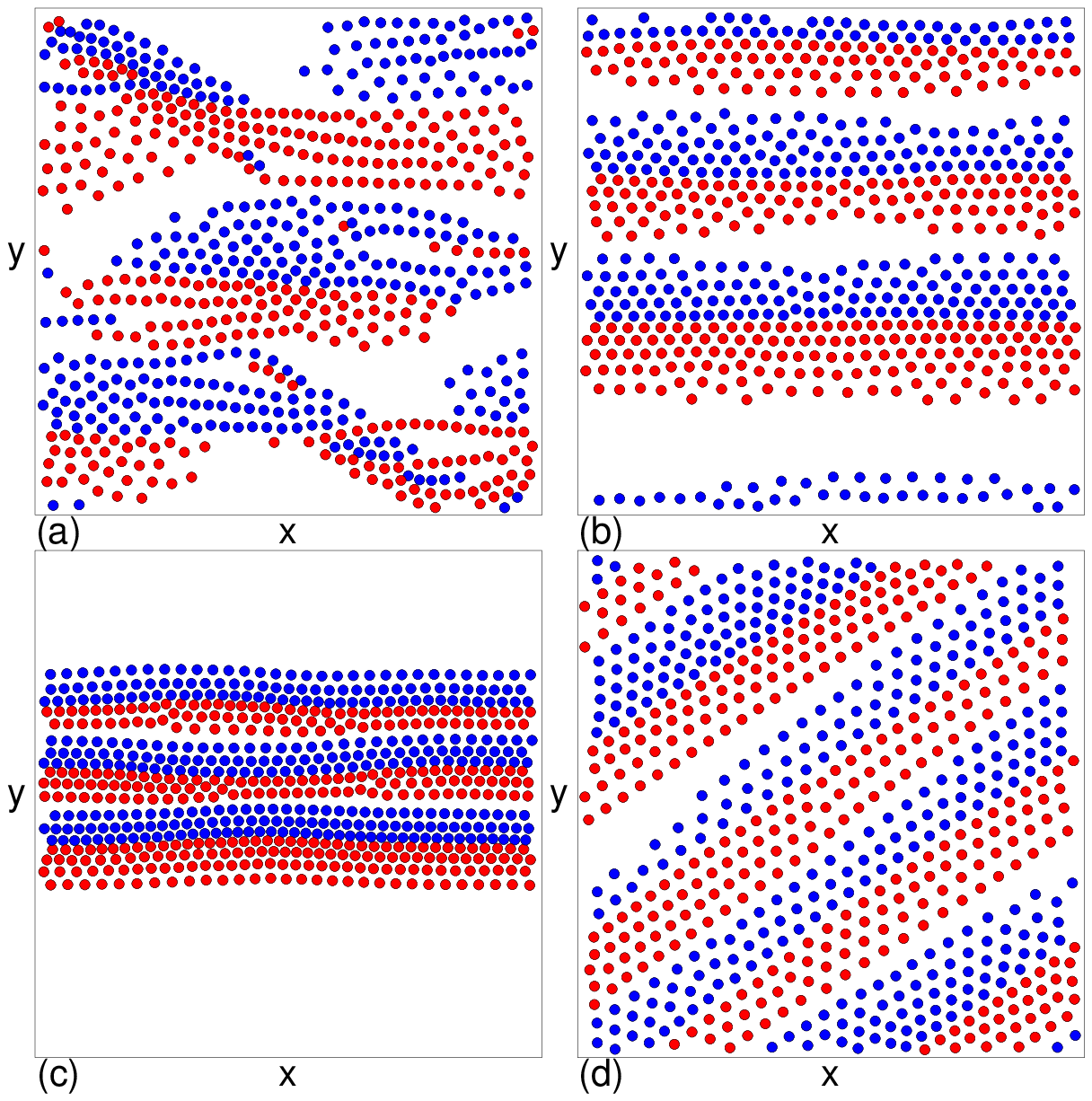}
\caption {Snapshots of particle positions for species A (blue) and
B (red) for the system from Fig.~\ref{fig:1} with
$\omega=2\times 10^{-6}$, $A=1.0$, and constant B species drive
$F_D^B=1.0$.
(a) A partially disordered state at $t/\tau=0.53$.
(b) A compressed lane state at $t/\tau=0.55$.
(c) A highly compressed lane state at $t/\tau=0.65$.
(d) A tilted lane state at $t/\tau=0.73$.
}
\label{fig:3}
\end{figure}

Near $t/\tau=0.53$ in Fig.~\ref{fig:1}
there is another jump in $V^{B}_{x}$, which reaches a value
close to $V^B_x=1.0$ for $0.53 \leq t/\tau \leq 0.67$. Over this same
interval of time,
$V^{A}_{y}$ switches from positive to negative values.
In Fig.~\ref{fig:3}(a) we illustrate the particle configurations
just after $t/\tau=0.53$
where the system is temporarily disordered and
is in the process of forming a laned state aligned
along the $x$ direction.
Although this newly formed lane state is aligned in $x$,
the drive on species A is now increasing in the 
$-y$ direction and decreasing in the $-x$ direction,
so the species A particles begin to press against
the species B particles and the stripe state becomes compressed
as shown in Fig.~\ref{fig:3}(b) at $t/\tau=0.55$.
The lanes become increasingly compressed, as indicated
in Fig.~\ref{fig:3}(c) at $t/\tau=0.65$,
just before the system switches to the new uniform tilted lane state
shown in Fig.~\ref{fig:3}(d) at $t/\tau=0.73$.
This tilted lane state becomes increasingly compressed
in turn before
another switching event occurs.
Since the initial starting state
was fully mixed,
the first locked state from
$t/\tau=0$ to $t/\tau=0.14$
does not exactly repeat at the start of the
second ac drive cycle,
but the other locked phases and switching
events do repeat for every cycle.
The switching transitions occur at nearly the same time during
each drive cycle,
but there can be slight variations in the exact switching
time from one ac cycle to the next.

We note that there is an asymmetry in $V^B_y$ in
Fig.~\ref{fig:1}(b). Despite the fact that $F^B_D$ is zero along the
$y$ direction, the species B particles do not move equally
along
both the $+y$ and $-y$ directions; instead, $V^B_y$ is predominantly
negative or zero. Additionally, changes in $V^B_x$ are nearly always
in the direction of reducing $V^B_x$ from its drive-dominated value of
$V^B_x=1.0$, rather than being symmetrically distributed both above
and below $V^B_x=1.0$.
This is due to a chiral bias generated by
the finite $x$ direction drive applied to species B.
When the dc driven species B
interacts with the rotating species A,
species B develops a net Hall angle.
Over the course of one ac drive cycle,
species B on average moves in the negative $y$ 
direction with $\langle V^{B}_{y}\rangle = -0.1$.
It has a reduced velocity of
$\langle V^{B}_{x}\rangle=0.95$ along
the positive $x$ direction compared to what would be produced by
$F_D^B$ alone, due to a drag effect exerted by species A.
Species B therefore exhibits 
a Hall angle of $\theta_H=-5.7^{\circ}$.
The value of the Hall angle for a dc driven
non-chiral particle moving through a medium of spinning particles
depends on the frequency of the circular motion and on the
drive applied to the non-chiral particles \cite{Reichhardt19a}.

\begin{figure}
\includegraphics[width=\columnwidth]{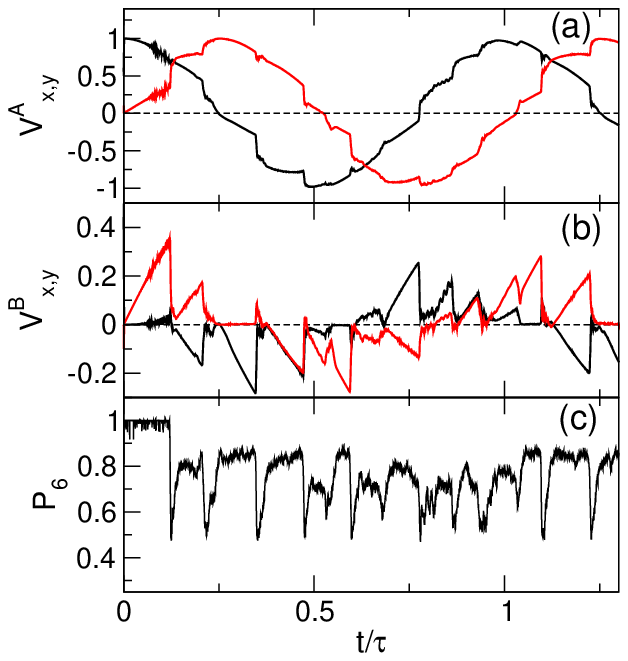}
\caption{Velocity and topological order for a system
where the drive on species A is rotated with $\omega=1\times 10^{-6}$ and
$A=1.0$,
but species B is passive with $F^{B}_{D} = 0.0$.
(a) $V^{A}_{x}$ (black) and $V^A_y$ (red) vs $t/\tau$.
(b) $V^{B}_{x}$ (black) and $V^B_y$ (red) vs $t/\tau$.
(c) $P_{6}$ vs $t/\tau$.
The finite velocity of species B is produced entirely by
interactions with species A.
}
\label{fig:4}
\end{figure}

\begin{figure}
\includegraphics[width=\columnwidth]{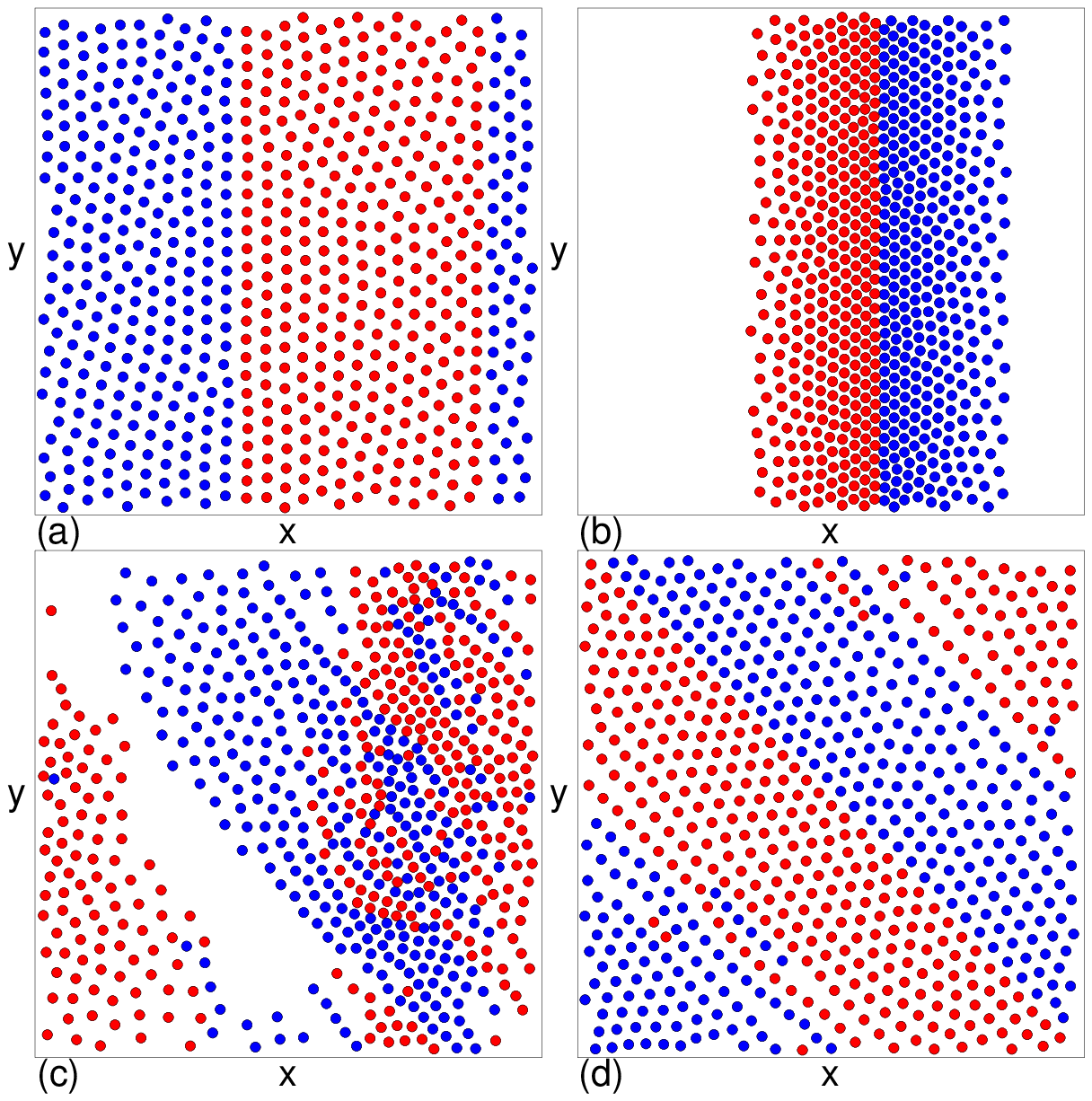}
\caption {Snapshots of particle positions for species A (blue) and B (red)
for the system from Fig.~\ref{fig:4} with
$\omega=1\times 10^{-6}$ and $A=1.0$, where species B is passive with
$F_D^B=0.0$.
(a) At $t/\tau=0.252$ of the cycle, there is
a uniform jammed state where species B is moving in the $-x$ direction
while being pushed by species A.
(b) At $t/\tau=0.32$ the stripe is more compressed.
(c) At $t/\tau=0.35$ the stripe breaks apart.
(d) At $t/\tau=0.375$ there is a uniformly tilted lane state.
This motion is illustrated in the Supplemental Movie.
}
\label{fig:5}
\end{figure}

\begin{figure}
\includegraphics[width=\columnwidth]{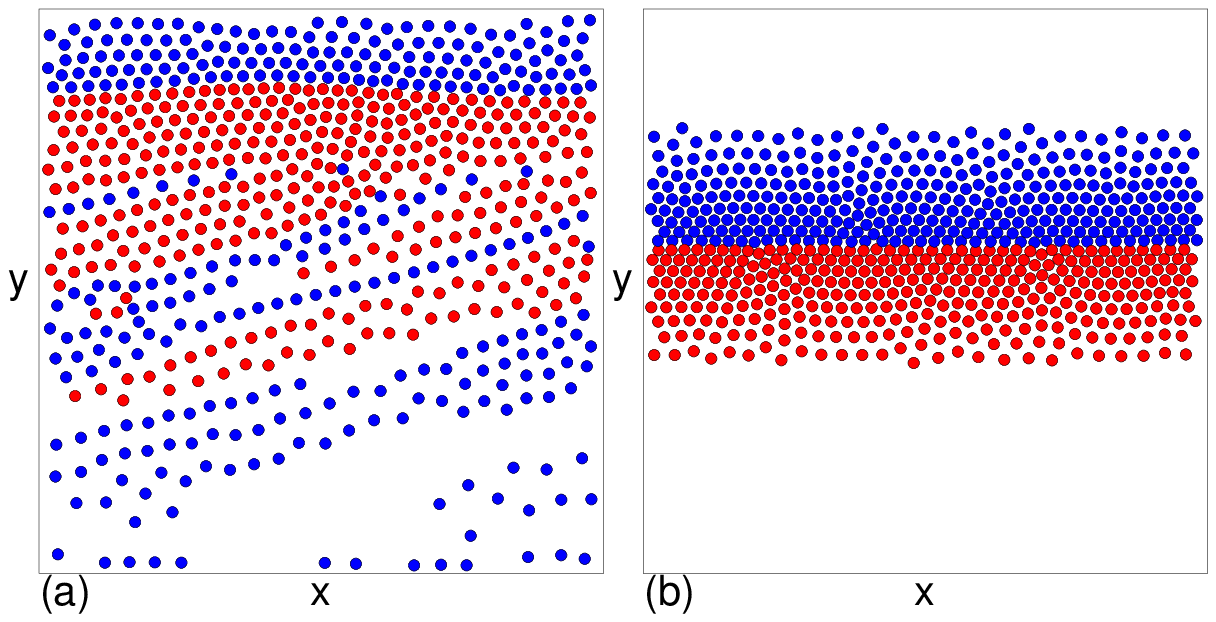}
\caption {Snapshots of particle positions for species A (blue) and B (red)
for the system from Fig.~\ref{fig:4} with
$\omega=1\times 10^{-6}$ and $A=1.0$, where species B is passive with
$F_D^B=0.0$.
(a) Transitional state at $t/\tau=0.536$.
(b) Fully developed clogged state at $t/\tau=0.58$. The
particles are locked along $y$ but translating along $-x$.
This motion is illustrated in the Supplemental Movie.
	}
\label{fig:6}
\end{figure}

We can eliminate the Hall effect by turning off the drive on species B.
In Fig.~\ref{fig:4}(a) we plot $V^{A}_{x}$ and $V^{A}_{y}$ versus $t/\tau$
for the same system from Fig.~\ref{fig:1} with
$A=1.0$ but
at $\omega=1 \times 10^{-6}$ and $F^{B}_{D}=0.0$, so that
the sample contains
actively rotating 
species A particles and passive species B particles.
Figure~\ref{fig:4}(b) shows
the corresponding $V^{B}_{x}$ and $V^{B}_{y}$ versus $t/\tau$,
and in Fig.~\ref{fig:4}(c) we plot $P_{6}$ versus $t/\tau$.
We find a set of laning phases
similar to those in the finite $F^B_D$ system described above,
but the number of laning states is
larger due to the lower driving frequency,
and the velocity components $V^B_x$ and $V^B_y$ of
species B are now symmetric about zero due to the lack of a Hall angle.
The motion of species B is produced solely by interactions with
the driven species A.
Here it is easier to see
that there are phases in 
which the species B motion occurs along a
$45^{\circ}$ angle, with $|V^{B}_{x}|$ and $|V^{B}_{y}|$ 
locked together.
At around $t/\tau=0.255$
there is a $y$-aligned stripe state, illustrated in
Fig.~\ref{fig:5}(a),
where species B is being pushed in 
the $-x$ direction by species A,
while species A moves primarily in the $-x$ direction with a 
smaller component
of motion along the $+y$ direction.
As time advances,
this lane or stripe becomes increasingly compressed,
as shown in Fig.~\ref{fig:5}(b) at $t/\tau=0.32$,
breaks up at the transition point $t/\tau=0.35$, as illustrated in 
Fig.~\ref{fig:5}(c), and is replaced
by the new tilted uniform state shown in Fig.~\ref{fig:5}(d)
at $t/\tau=0.375$.
Figure~\ref{fig:6}(a) shows
the transition at $t/\tau=0.536$ to a clogged
stripe state that is motionless in the $y$ direction but
translating along $-x$,
while
Fig.~\ref{fig:6}(b) shows the fully developed clogged state
at $t/\tau=0.58$. In this clogged state, species A
is moving much more rapidly along $-x$ than species B, so there is
slip occurring along the A-B boundary.

\begin{figure}
\includegraphics[width=\columnwidth]{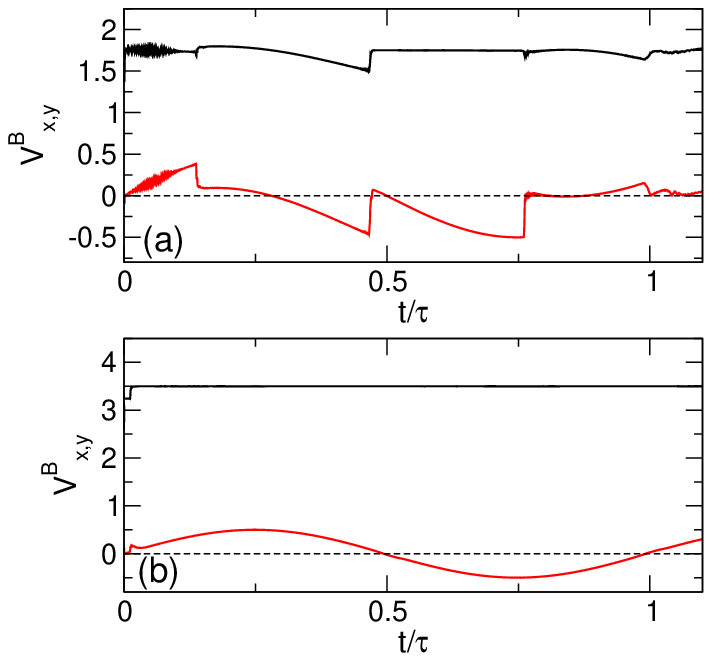}
\caption{Velocity and topological order
for the system from Fig.~\ref{fig:1}(a) with
$\omega=2\times 10^{-6}$ and $A=1.0$, but where
the constant drive $F_D^B$ of species B in the $x$-direction is changed.
(a) $V^{B}_{x}$ (black) and $V^B_y$ (red) vs $t/\tau$ at
$F^{B}_{D} = 1.75$, showing a reduced number of switching events.
(b) The same but for $F^{B}_{D} = 4.0$,
where there are no switching events and
the system forms a single lane aligned along the $x$ direction.
}
\label{fig:7}
\end{figure}

The number of switching events is affected by the value
of $F^{B}_{D}$, and is maximal
for $F^{B}_{D} = 0.0$.
In Fig.~\ref{fig:7}(a) we
plot $V^{B}_{x}$ and $V^{B}_{y}$ versus $t/\tau$
for the same system from Fig.~\ref{fig:1}
but at $F^{B}_{x} = 1.75$, where the number of
switching events per ac drive period is reduced.
In Fig.~\ref{fig:7}(b) we plot the same quantities for the same
system but with $F^B_D$ increased to $F^B_D=4.0$.
Here, no switching events occur at all, and
the system forms
a single lane aligned with the $x$ direction.
The fact that $V^{B}_x=4.0$, the same value as $F^B_D$, indicates that
there is no dragging effect from species A on species B,
and the lack of drag is also reflected in the fact that
$V^{B}_{y}$ undergoes smooth sinusoidal oscillations without jumps.
It is still possible for the single 
lane to change width periodically as
species A exerts forces in the positive or negative $y$ direction on
species B, but the lane always remains aligned in the $x$ direction.

\begin{figure}
\includegraphics[width=\columnwidth]{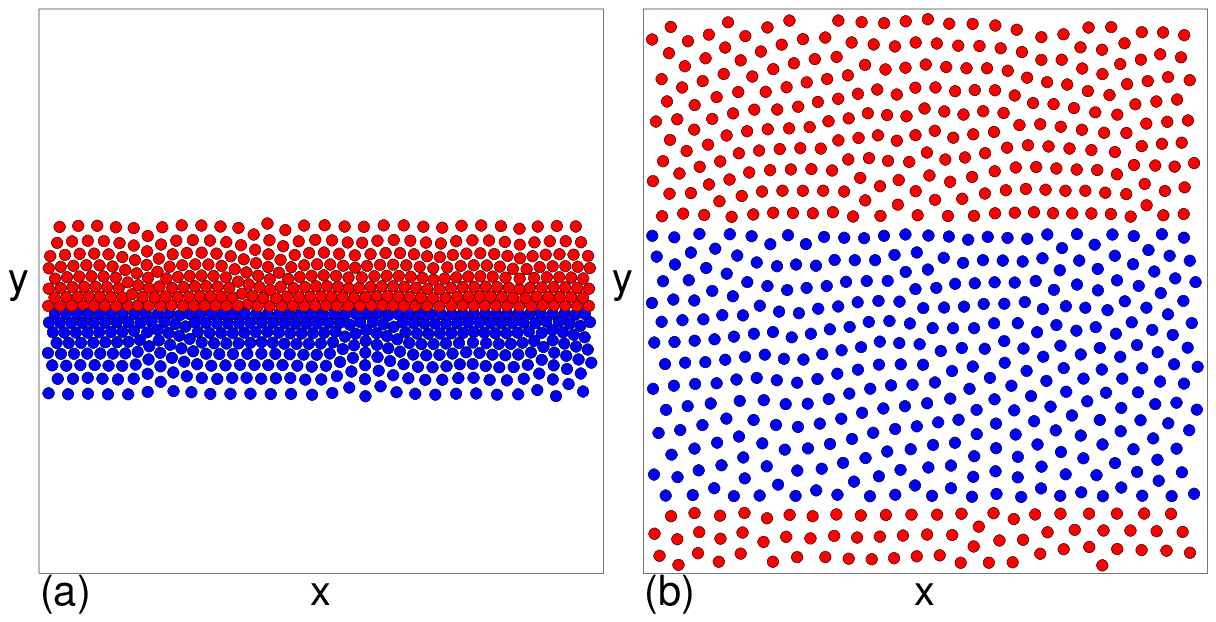}
\caption{Snapshots of particle positions for species A (blue) and B (red)
for the system from Fig.~\ref{fig:7}(b) with
$\omega=2\times 10^{-6}$, $A=1.0$, and a constant
B species drive of $F_D^B=4.0$.
(a) A compressed lane state at $t/\tau=0.25$, with only a single lane present.
(b) A uniform lane state at $t/\tau=0.5$.
}
\label{fig:8}
\end{figure}

In Fig.~\ref{fig:8}(a) we show an image of the particle positions
in the laned state for the $F_D^B=4.0$
system from Fig.~\ref{fig:7}(b) at $t/\tau=0.25$ when the
positive $y$-direction drive on species A has reached its maximum value,
causing a strong compression of the lane. Here the entire lane structure
is translating as a unit in the $+y$ direction, while the two particle
species shear past each other in the $+x$ direction. 
At $t/\tau=0.5$ in Fig.~\ref{fig:8}(b),
the driving force in the $y$-direction on species A is zero,
and the system forms a uniform lane state that is stationary in the
$y$ direction but continuing to shear along the $+x$ direction.
The compressed lane state in
Fig.~\ref{fig:8}(a) exhibits a clear density gradient, with low particle
density far from the A-B interface and high particle density close to
the interface.
This density variation, combined with the intermediate range of the
interaction potential, results in the formation of an arc-like conformal lattice
structure on each side of the A-B boundary.
Conformal lattices have been studied in two-dimensional systems
for particles with softer repulsive interactions
that are pushed against a wall by a force such as gravity \cite{Rothen96}.
For other systems such as superconducting vortices and
magnetic skyrmions, conformal lattice structures have
also been observed when the particles
are pushed against a wall or a barrier \cite{Menezes17,Souza23}.
In Fig.~\ref{fig:8}(a) the
conformal lattice forms when the two particle
species push against the A-B interface.

For oppositely driven hard disks, similar compressed lane structures
can appear, but since the disk-disk interactions are very short range,
the particles are in direct contact with each other and form a jammed
lane structure in which no density gradient is present
\cite{Reichhardt18}.
For the Yukawa interacting particles considered here,
when the particles are driven in
strictly opposite directions, the interactions are
sufficiently soft that the particles are able to slip past
each other, so we do not observe the formation of jammed states or
gradient states of the type shown in Fig.~\ref{fig:8}(b).
For the combination of dc and ac driving as in
Fig.~\ref{fig:8}, it is very difficult for A and B particles to slip
past each other because not only is there a density gradient at the interface,
but also each species is sliding past the other species along the $x$
direction, so that by the time a particle is prepared
to move into an interstitial
site across the interface, that site has already traveled away to a new
location.
In general, for $F^{B}_D/A > 2.5$
the system only forms a single lane that is
aligned in the direction of the dc drive on species B, which in this case
is the $x$ direction.

\section{Effect of Changing Circular Drive Frequencies}

\begin{figure}
\includegraphics[width=\columnwidth]{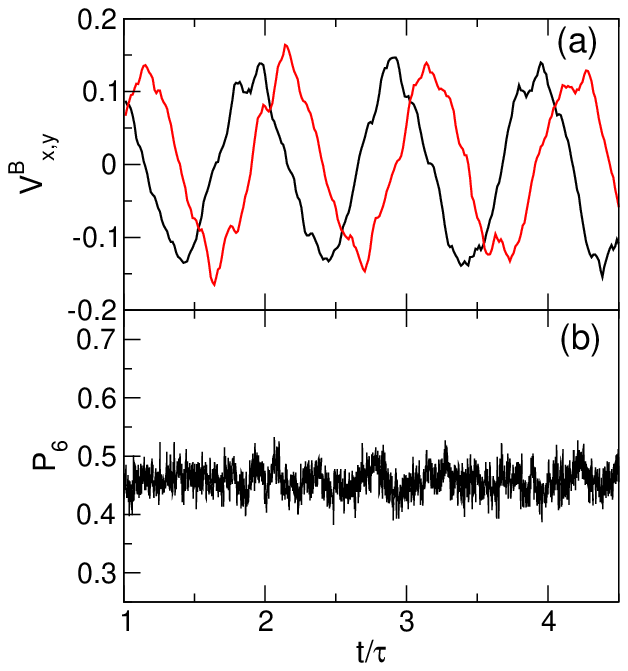}
\caption {(a) $V_x^B$ (black) and $V_y^B$ (red) vs $t/\tau$
for a system with
$A=1.0$, passive B species with
$F^{B}_{D} = 0.0$, and $\omega = 2\times 10^{-4}$, a frequency
that is
100 times higher than the system in Fig.~\ref{fig:4}.
(b) The corresponding $P_{6}$ vs $t/\tau$.
}
\label{fig:9}
\end{figure}

We next show that in addition to its dependence on the value
of $F^{B}_{D}$, the laning and switching behavior is also strongly
affected by the value of the ac drive frequency $\omega$.
In Fig.~\ref{fig:9}(a) we plot
$V^{B}_{x}$ and $V^B_y$ versus $t/\tau$ for a system with
$F^{B}_{D} =0.0$ or
passive B particles
at $\omega = 1 \times 10^{-4}$,
a frequency 100 times higher than that of the system in
Fig.~\ref{fig:4}.
For this rapid variation of the drive,
the sharp switching events are lost and the net velocity
$V^B_x$ and $V^B_y$ of species B is significantly reduced,
indicating a decreased drag effect by species A
compared to the lower frequency system in Fig.~\ref{fig:4}.
In the plot of the corresponding $P_6$ versus $t/\tau$ shown in
Fig.~\ref{fig:9}(b), the sharp jumps up and down that occurred at
lower frequencies during switching events have vanished,
and the system is always heavily disordered.

\begin{figure}
\includegraphics[width=\columnwidth]{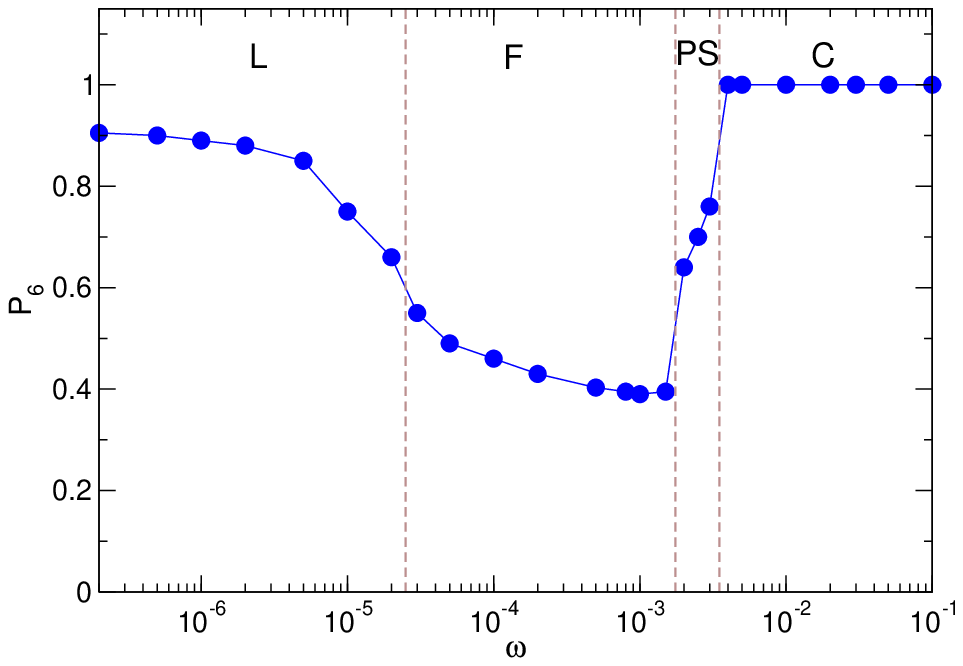}
\caption {$P_{6}$ vs $\omega$ measured after one period in a system
with
$A=1.0$ and passive B species with $F_D^B=0.0$.
For low frequencies the system forms a laned state (L).
At intermediate frequencies we find a fluid (F),
followed by a phase-separated state (PS) and,
at the highest frequencies, a crystal (C).
}
\label{fig:10}
\end{figure}

In Fig.~\ref{fig:10} we plot $P_{6}$ measured after one ac drive period
versus drive frequency $\omega$. Here we report the largest value of $P_6$
that occurred during the drive period.
Ordered lane states appear at low frequencies of
$\omega < 1 \times 10^{-5}$.
For $1 \times 10^{-5} \leq \omega < 1.5 \times 10^{-3}$,
the system forms a disordered or fluid state during the entire ac drive
cycle. There is a narrow window of phase separated states at higher
frequencies followed by the formation of a crystal at the highest
frequencies.

\begin{figure}
\includegraphics[width=\columnwidth]{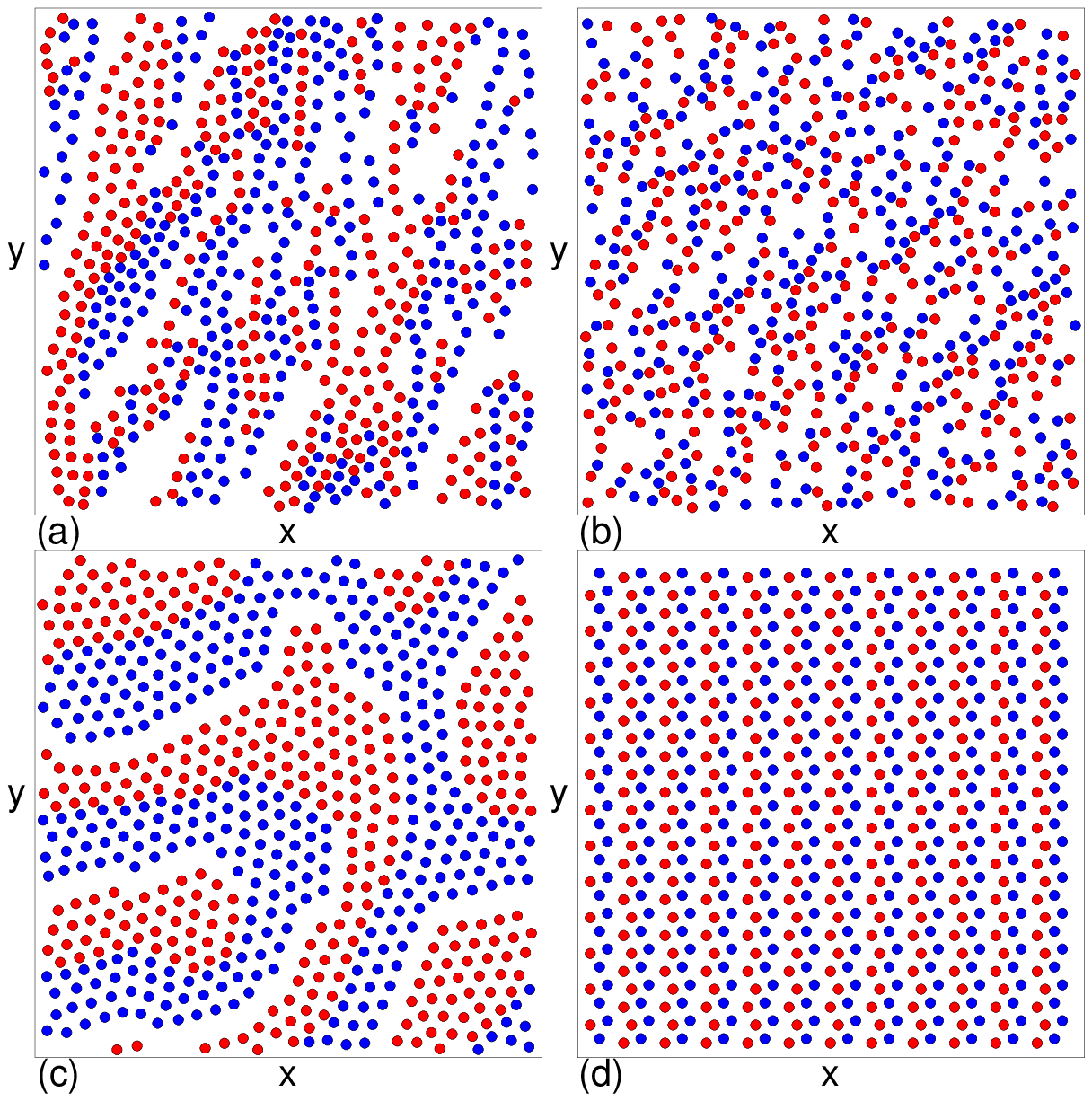}
\caption {Snapshots of particle positions for species A (blue) and B (red)
for the system from Fig.~\ref{fig:10} with
$A=1.0$ and
passive B species with $F_D^B=0.0$.  
(a) Disordered lane state at $\omega = 1 \times 10^{-4}$ from the same
system illustrated in Fig.~\ref{fig:9}.
(b) $\omega = 0.001$, where the system is maximally disordered.
(c) A phase-separated state at $\omega = 0.003$.
(d) A crystal state at $\omega = 0.01$.
}
\label{fig:11}
\end{figure}

In Fig.~\ref{fig:11}(a) we show the particle configurations for the
system from Fig.~\ref{fig:9} in the disordered state
where there is some weak lane-like ordering present.
Figure~\ref{fig:11}(b) shows that
the same system at $\omega = 0.001$ is maximally disordered.
As $\omega$ increases,
the radius of the circular orbits executed by the
species A particles becomes smaller,
and the system enters a new phase separation regime
when the orbit drops below a size of
a few times the average distance between particles.
The phase separated state is illustrated
in Fig.~\ref{fig:11}(c) at $\omega = 0.003$.
At even higher $\omega$, when the circular orbit of species A
becomes smaller than the average spacing between particles,
the species A particles act more like passive point particles,
and the system forms a crystal due to the repulsion between the
particles, as shown in Fig.~\ref{fig:11}(d) at $\omega = 0.01$.

The phase-separated state shown in Fig.~\ref{fig:11}(c)
is similar to the phase separation
found for mixtures of rotating and passive disks \cite{Reichhardt19}.
In general, when phase separation occurs in 
a hard disk system of this type, the passive particles form a dense
jammed state,
and the phase separation is robust over
a wide range of ac drive frequencies \cite{Reichhardt19}.
The Yukawa particles considered in the present work have a softer
particle-particle interaction potential, making it easier for the particles
to slip past each other compared to a hard disk system. As a result,
phase separation occurs only for relatively small orbit radii.
For large circular drive amplitudes in the hard disk system,
the particles form a fluid;
however, in previous work on hard disks
the low-frequency limit was not explored, so it is not known whether
the hard disks would form
lane-like states similar to what we
find here for the Yukawa particles.

\begin{figure}
\includegraphics[width=\columnwidth]{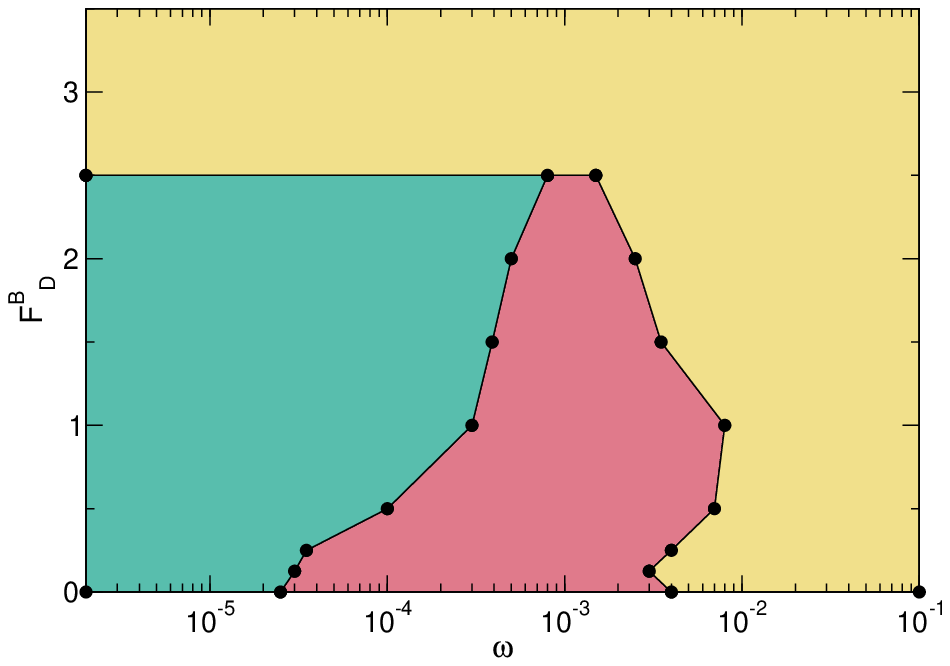}
\caption{Dynamic phase diagram as a function of $F_D^B$ vs $\omega$
for the system from Figs.~\ref{fig:9} and \ref{fig:10} with
$A=1.0$.
Green: the switching laned state.
Pink: the fluid state.
Yellow: a single lane aligned in the $x$ direction.
The phase separated state
and crystal state that form for $F^{B}_{D} = 0.0$ are not shown.
}
\label{fig:12}
\end{figure}

In Fig.~\ref{fig:12} we plot a dynamic phase diagram as a function of
$F_D^B$ versus $\omega$ for the system from
Figs.~\ref{fig:9} and \ref{fig:10}.
Highlighted on the diagram are
the laned state, the fluid state,
and a state with a single lane that is aligned in the $x$ direction.
We note that for
$F^B_D=0.0$, additional phase-separated and crystal
states are present that are not shown in the diagram but will
be discussed elsewhere.
When $F^B_D \leq 2.5$, we observe all three phases, while for large
$F^B_D$, there is only
the single lane state aligned in $x$,
even for low ac drive frequencies.
This result indicates
that the switching between different laned states is robust at
low $\omega$ for lower values of $F^B_D$.
The single lane state
that appears for high $\omega$ at
finite $F^{B}_{D}$ is associated with a phase separation in the system.
The finite driving of the species B particles permits each species
to form wide lanes that are aligned with the $x$ direction, which is the
direction in which species B is driven.
For high $\omega$ and low $F^{B}_{D}$,
the lane state becomes more disordered,
but some alignment along the $x$ direction persists.
Thus it may be possible to subdivide the response into a larger number
of dynamical states beyond those illustrated in
Fig.~\ref{fig:12}. The phase diagram captures the most important
feature, however, which is the regime in which
switching between the different lane states can occur.

\section{Thermal and Melting Effects}

\begin{figure}
\includegraphics[width=\columnwidth]{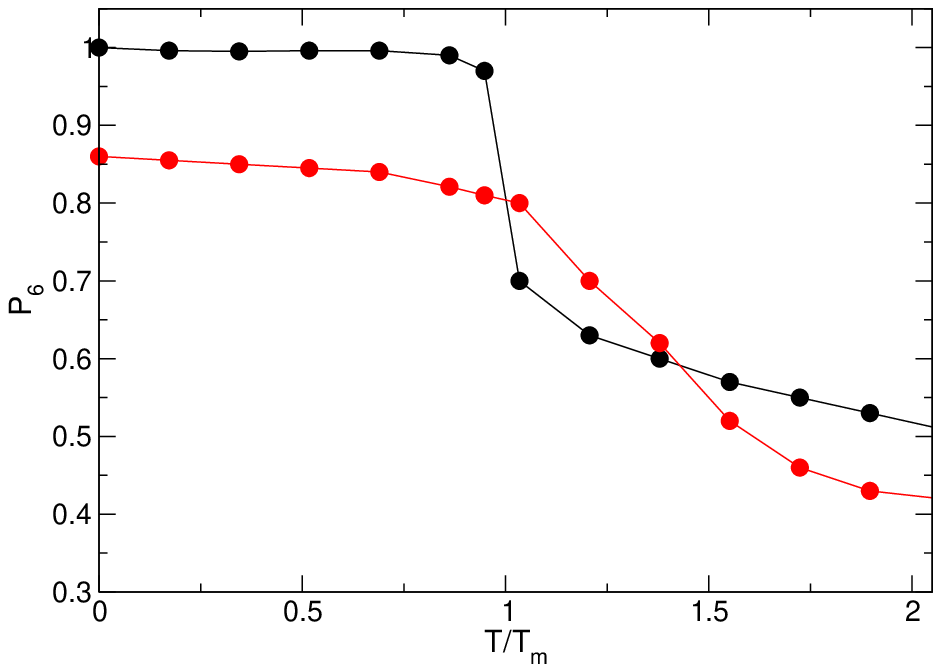}
\caption {$P_{6}$ vs
temperature $T/T_m$ for a system
under either no driving on either species (black)
or with $F_D^B=0$ passive species B
and an ac circular drive on species $A$ with amplitude $A=1.0$ and
frequency $\omega = 1\times10^{-6}$ (red).
Here $T_m$ is the temperature at which
a proliferation of topological defects
occurs for the non-driven or equilibrium system.
}
\label{fig:13}
\end{figure}

To test the robustness of the laning and switching states, we introduce
finite temperature to the simulations.
We add Langevin kicks with the properties
$\langle F_T\rangle=0$ and
$\langle F_T^i(t) F_T^j(t^\prime)\rangle= 2\eta k_B T \delta_{ij}\delta(t-t^\prime)$,
where $k_B$ is the Boltzmann
constant.
In Fig.~\ref{fig:13} we plot $P_{6}$
versus $T/T_m$ for a system
with no driving on either species.
This corresponds to the equilibrium thermal melting of a two-dimensional
Yukawa crystal, where $T_m$ is the temperature at which
a proliferation of topological defects
occurs in the equilibrium system.
For $T/T_m < 1.0$, the system is crystalline with
$P_{6}$ close to $1.0$,
and for $T/T_m \geq 1.0$ there is
a drop off in $P_6$.
Also plotted is $P_6$ versus $T/T_m$ for a system with passive
or $F_D^B=0.0$
species
B particles and ac driven
species A particles with amplitude $A=1.0$ and
frequency $\omega = 1\times10^{-6}$.
The value of $P_6$ is taken from the laning portion of the ac drive
cycle at which the system is the most ordered.
For this low ac driving frequency,
at $T/T_m = 0.0$ the system forms the well defined lanes described
in the previous sections.
As $T/T_m$ increases from zero, the value of $P_6$ for the driven system
drops below its value in the equilibrium system
due to the fact that the edges of the lanes are decorated by
defects,
but the system still remains ordered.
There is a temperature interval
$1.0 < T/T_m < 1.5$ over which
the driven system exhibits a greater amount of order, indicated by
a higher value of $P_6$,
than the equilibrium system,
but for $T/T_m > 1.5$,
the driven system is more disordered than the equilibrium system.
The value of $P_6$ for
the driven system remains relatively constant up to $T/T_m = 1.0$,
and in this temperature range
the system exhibits the same types of laning and switching patterns
found for $T/T_m=0$,
indicating that these states remain robust against the introduction of
a finite temperature.
An interesting effect is that
even for $1.0 < T/T_m < 1.5$, the
driven system still contains lanes that have
some triangular ordering.
This occurs due to the fact that the ac drive compresses the lanes during
a portion of the drive cycle. The melting temperature of
repulsive Yukawa particles depends on the density of the system and
increases with increasing density. Thus, when the lanes are compressed
and the temperature is not yet too far above $T_m$,
the
compressed lane can drop below the equilibrium melting temperature for
the
compressed density. As a result the lane tends to recrystallize.

\begin{figure}
\includegraphics[width=\columnwidth]{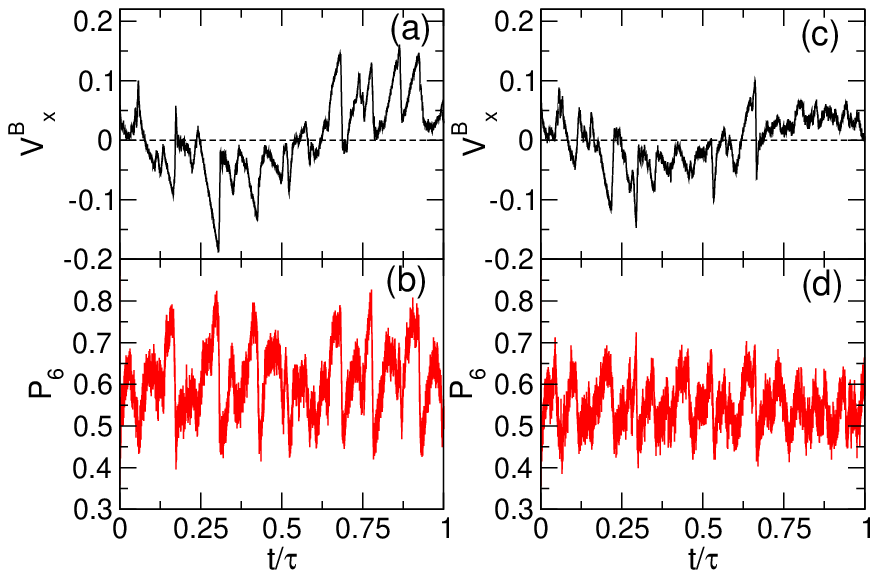}
\caption {(a) $V^B_{x}$ and (b) $P_{6}$ vs $t/\tau$ for
the ac driven system with 
$F_B=0$, $A=1.0$, and $\omega=1 \times 10^{-6}$ from
Fig.~\ref{fig:13} at $T/T_m=0.68$. Here there are clear jumps associated with
switching among different laning states.
(c) $V^B_x$ and (d) $P_6$ vs $t/\tau$ for the same system at
$T/T_m = 1.2$, where switching between lane states still occurs but
the amount of fluctuations has increased.
}
\label{fig:14}
\end{figure}

\begin{figure}
\includegraphics[width=\columnwidth]{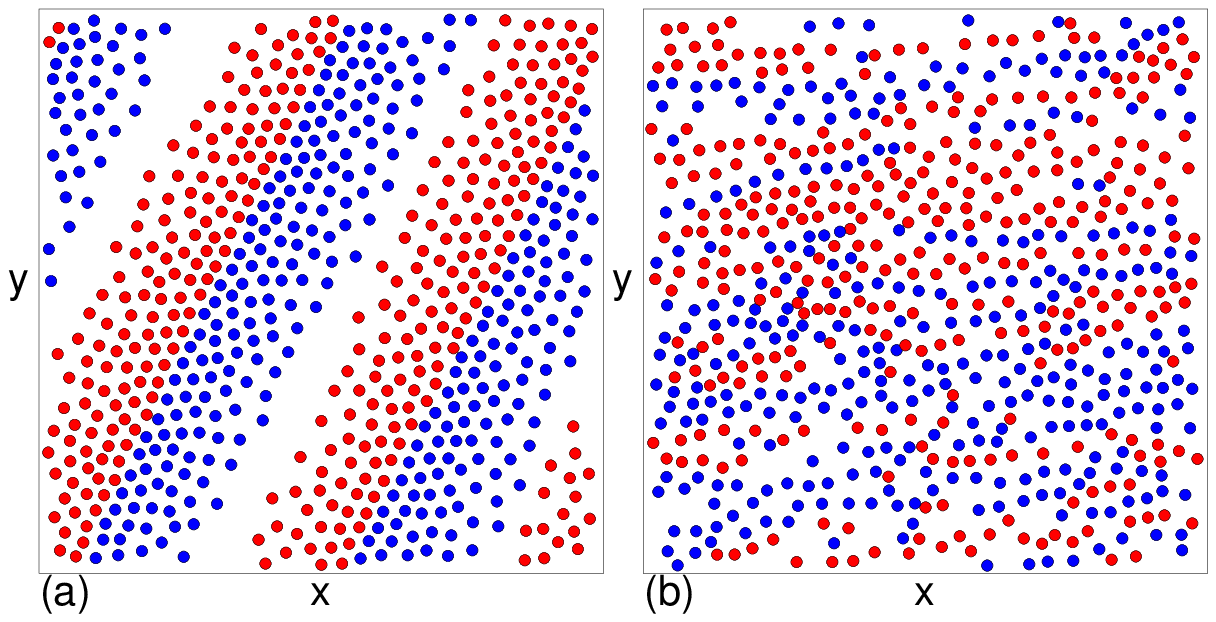}
\caption {Snapshots of particle positions for species A (blue) and B (red)
for a system 
with
$F^B_D=0$, $A=1.0$, and $\omega=1 \times 10^{-6}$.
(a) Laning state
for the system from Fig.~\ref{fig:14}(c,d) 
at $T/T_m = 1.2$.
(b) Fluid phase for the system in
Fig.~\ref{fig:15}(a,b) at $T/T_m = 1.72$.}
\label{fig:16}
\end{figure}

\begin{figure}
\includegraphics[width=\columnwidth]{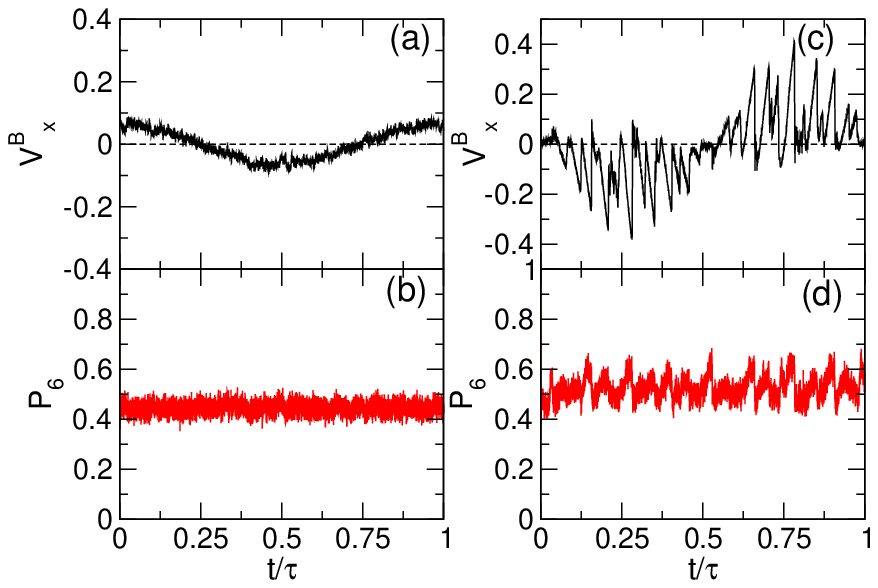}
\caption {(a) $V_x^B$ and (b) $P_6$ vs $t/\tau$ for the ac driven system
with
$F_B=0$, $A=1.0$, and $\omega=1\times 10^{-6}$ from
Fig.~\ref{fig:13} at $T/T_m=1.72$.
Here the switching and laning behaviors are lost.
(c) $V_x^B$ and (d) $P_6$ vs $t/\tau$ for the same system at the same
temperature of $T/T_m = 1.72$, but for
an ac drive amplitude of $A=4.0$.
The switching behavior is restored.
}
\label{fig:15}
\end{figure}

In Fig.~\ref{fig:14}(a,b) we
plot $V^B_{x}$ and $P_{6}$ versus $t/\tau$ for the
ac driven system from Fig.~\ref{fig:13}
at $T/T_m = 0.68$.
The same jumps in the velocity and dips and peaks in $P_6$ appear as were
found for $T/T_m = 0.0$.
At $T/T_m=1.2$ in the same system,
the plots of $V^B_x$ and $P_6$ versus $t/\tau$ in
Fig.~\ref{fig:14}(c,d)
indicate that there is still a strong signature of
laning in the dips and peaks in $P_{6}$.
The maximum values reached by $P_6$ are lower than what appears at
lower temperatures, and the overall level of fluctuations is increased.
In Fig.~\ref{fig:16}(a) we show
a snapshot of the particle positions for the
$T/T_m=1.2$ system from Fig.~\ref{fig:14}(c,d).
The lane structure clearly persists at this finite temperature
thanks to the compression of the lanes by the ac drive,
and the densest portion of each lane retains the greatest amount of
triangular ordering, consistent with the increase in melting temperature
for a denser Yukawa system.
In Fig.~\ref{fig:15}(a,b),
we plot $V^B_{x}$ and $P_{6}$ versus $t/\tau$ for
the same system at $T/T_m = 1.72$.
Here $V^{B}_{x}$ varies smoothly and has an overall low value,
indicating a reduction in the extent to which species A particles can
drag the species B particles. At the same time,
$P_{6}$ contains neither distinct dips nor peaks,
but fluctuates around $P_6=0.46$, indicating
that the system is in a fluid phase as illustrated in Fig.~\ref{fig:16}(b).
If we increase the ac driving amplitude,
we can achieve further compression of the system
during each ac drive cycle, and the laned states can reemerge,
as shown in Fig.~\ref{fig:15}(c,d)
where we plot $V^B_{x}$ and $P_{6}$ versus $t/\tau$
at $T/T_m = 1.72$
but for a higher ac drive amplitude of
$A=4.0$.
Here there is clear evidence of laning in
the peaks and jumps in $V^{B}_{x}$ and
the local peaks in $P_{6}$.

\begin{figure}
\includegraphics[width=\columnwidth]{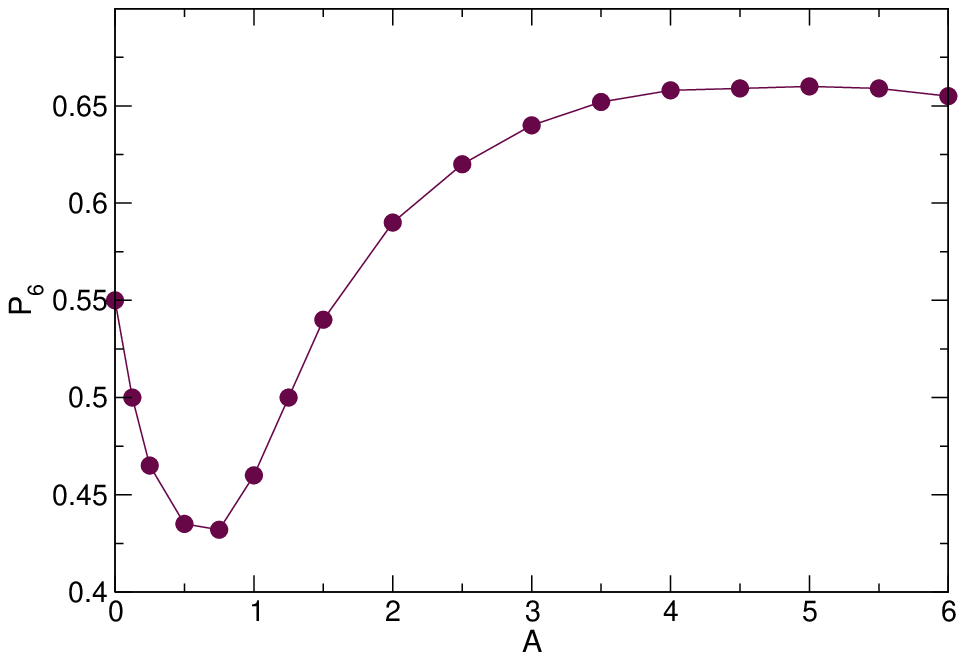}
\caption {The maximum value of $P_6$ during an ac drive cycle
vs ac drive amplitude $A$ in the lane state for the
system from Fig.~\ref{fig:15}(b) with
$F^B_D=0$,
$\omega=1\times 10^{-6}$, and $T/T_m=1.72$.
}
\label{fig:17}
\end{figure}

In Fig.~\ref{fig:17} we plot
the maximum value reached by $P_6$ during the full ac drive cycle versus
the ac drive amplitude $A$ for the
passive species B system from Fig.~\ref{fig:15}(b) at
$T/T_m=1.72$.
At $A = 0.0$ the system is in an equilibrium liquid state with $P_{6} = 0.55$.
As the ac drive amplitude increases, no laning is present and
the system becomes even more disordered, reaching its minimum value of
$P_6$ at $A=0.75$.
For higher values of $A$,
a lane structure gradually begins to emerge,
and for $A \geq 2.0$, when the
cyclic compression is
sufficiently large,
some crystallization emerges in the densest portion of the lane.
This result indicates that the laning and switching dynamics
remain robust even at higher temperatures
provided that the ac drive amplitude is large enough to compress
a portion of the lanes into a crystalline state.

\section{Both Species Under Circular Drives}

\begin{figure}
\includegraphics[width=\columnwidth]{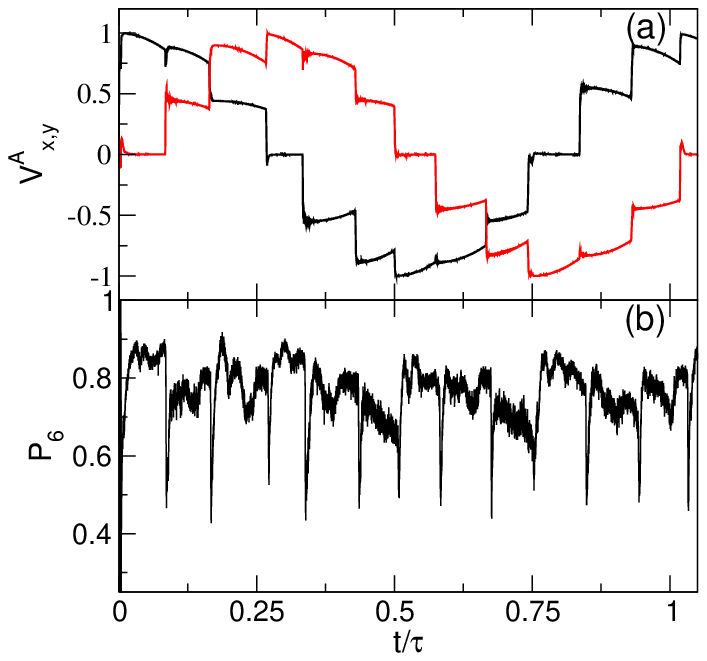}
\caption {(a) $V^A_x$ (black) and $V^A_y$ (red) vs $t/\tau$ for a system
where both species A and B are subjected to a circular
drive with $A=1.0$ and $\omega=1\times 10^{-6}$. The drive on species B
is out of phase by $\pi$ from the drive on species A.
(b) The corresponding $P_6$ vs $t/\tau$.
}
\label{fig:18}
\end{figure}

\begin{figure}
\includegraphics[width=\columnwidth]{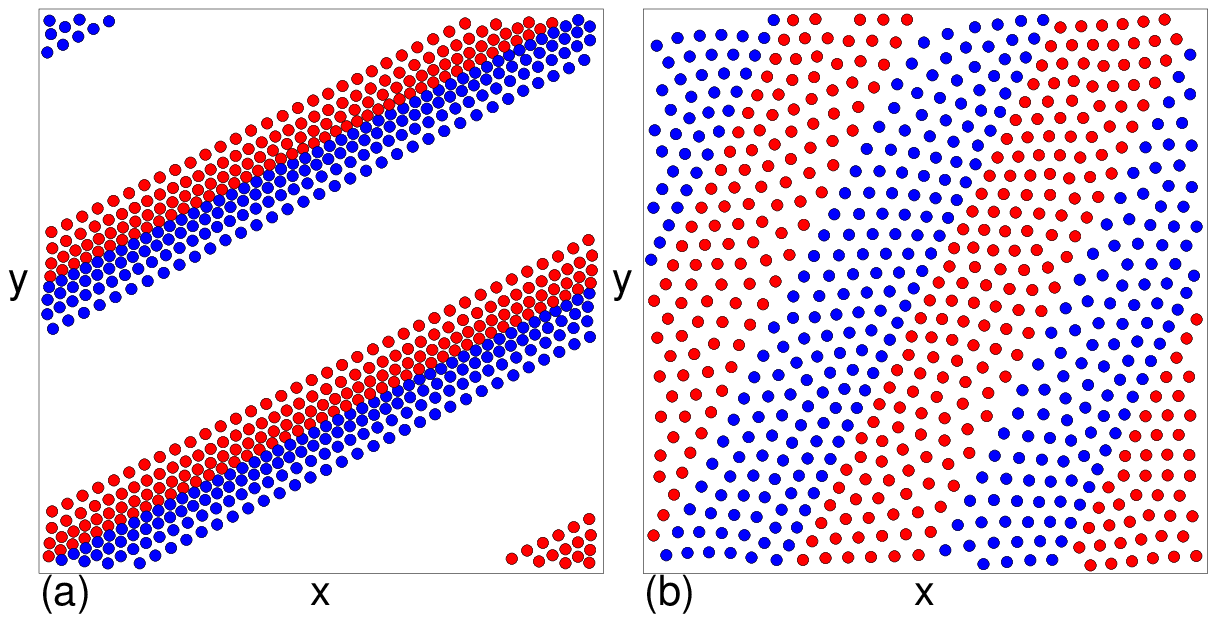}
\caption{Snapshots of particle positions for species A (blue)
and B (red) for the system from Fig.~\ref{fig:18}
in
which both species A and B are subjected to a circular drive with
$A=1.0$ and $\omega=1\times 10^{-6}$. The drive on species B is out
of phase by $\pi$ from the drive on species A.
(a) Compressed lanes just before the
second switching event at $t/\tau=0.164$.
(b) Uniform lanes just after the second switching event.}
\label{fig:19}
\end{figure}

We next consider the case in which both species A and species B are
subjected to an 
ac circular drive with the same chirality but with a phase shift of
$\pi$ with respect to each other.
We first examine a
low frequency drive with $\omega=1\times 10^{-6}$
at an ac drive amplitude of $A=1.0$.
In Fig.~\ref{fig:18}(a) we plot $V^{A}_{x}$ and $V^A_y$ versus
$t/\tau$, where we find a series of sharp jumps up and down.
The corresponding plot of $P_6$ versus $t/\tau$ in
Fig.~\ref{fig:18}(b) contains dips that coincide with the sudden
velocity changes.
For these driving conditions, the system forms lanes similar to those
observed when the drive on species B is a constant dc force rather
than an ac force.
The behavior of $V^B_x$ and $V^B_y$ versus $t/\tau$ (not shown) is the
same as that of $V^A_x$ and $V^A_y$ in Fig.~\ref{fig:18}(a).
The velocity switching signatures are
sharper than those that appear for
a constant drive on species B,
but the lane states show the same type of switching and
compression behavior.
In Fig.~\ref{fig:19}(a) we illustrate the
particle configurations just below the second switching event
at $t/\tau=0.164$, where the system forms compressed tilted stripes.
Figure~\ref{fig:19}(b) shows the system just above the second
switching event where a uniform tilted lane state appears.
We find similar behavior across the other switching events.
This result indicates that the laning and switching processes
should arise regardless of whether one or both species is subjected to
an ac circular drive, provided that the driving frequency is sufficiently
low. If both species are under ac circular driving, the drives must be
out of phase to produce the laning behavior.
If the rotational driving on both species is in phase,
the system does not show any switching
but instead smoothly rotates as a rigid triangular lattice.

\begin{figure}
\includegraphics[width=\columnwidth]{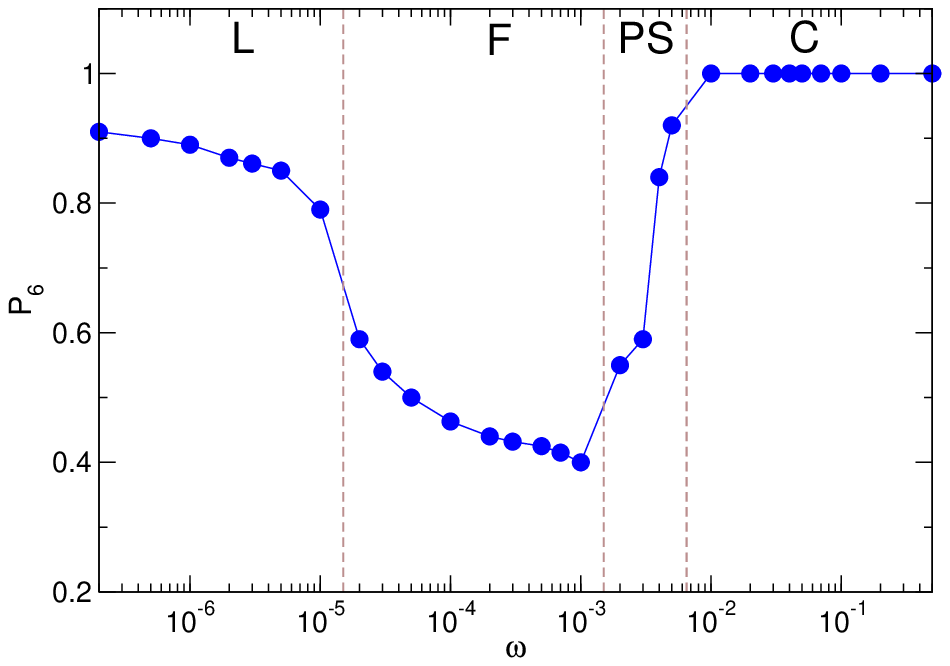}
\caption{The maximum value of $P_{6}$ during an ac drive cycle
vs ac drive frequency $\omega$
for the system in
Fig.~\ref{fig:18}
in which both species A and B are
subjected to a circular drive with $A=1.0$. The drive on species B is
out of phase by $\pi$ from the drive on species A.
We find a laned state (L) at low frequencies, a fluid state (F)
at intermediate frequencies, a phase separated state (PS)
at high frequencies, and a crystal (C) at the highest frequencies.        
}
\label{fig:20}
\end{figure}

We explore the effect of the driving frequency on a system where both particle
species are subjected to out of phase ac driving in
Fig.~\ref{fig:20},
where we plot $P_{6}$ versus
$\omega$ for the system from
Fig.~\ref{fig:18}.
Here the value of $P_6$ is measured
in the portion of the ac drive cycle during which lanes are forming,
which is when $P_6$ reaches its highest value.
We find a laned state at low frequencies, a fluid state
at intermediate frequencies, a phase separated state at high
frequencies, and a crystal at the highest frequencies.
These states are very similar
to what we observed in Fig.~\ref{fig:10}
where species B was passive rather than driven.
As in that case, it is necessary to apply high ac drive
frequencies to reach
the phase separated state and the crystal.
For high driving frequencies
when both species are under out of phase ac driving,
we observe several different types of crystals
that will be discussed in a separate work.

\begin{figure}
\includegraphics[width=\columnwidth]{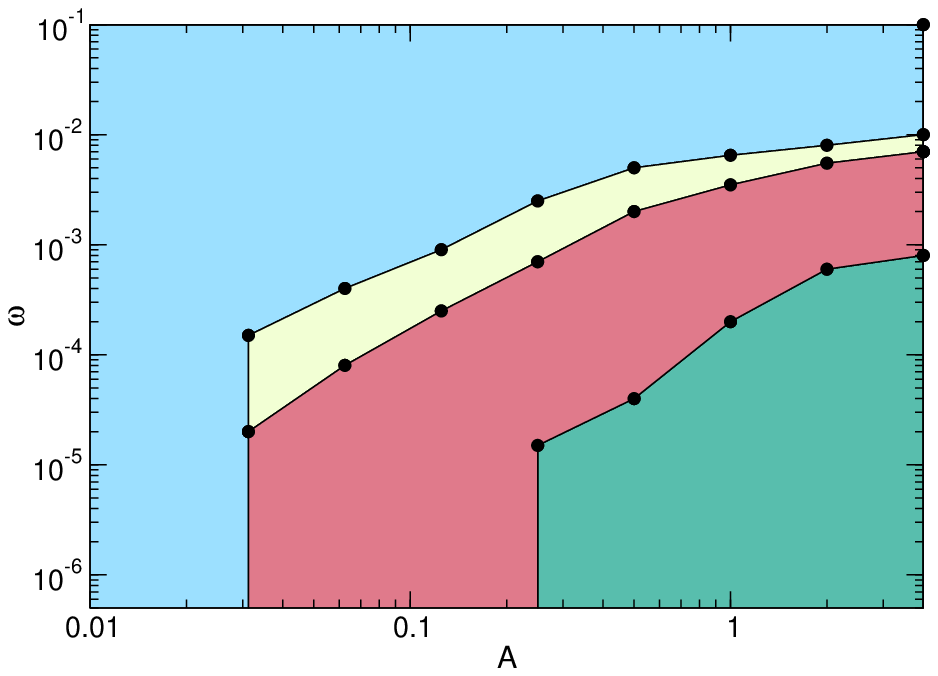}
\caption {Dynamic phase diagram as a function
of ac drive frequency $\omega$ versus ac drive amplitude $A$ for
the system in Fig.~\ref{fig:20}
where both
species A and B are subjected to a circular ac drive and in which
the drive on species B is out of phase by $\pi$ from the drive on
species A.
Green: the switching laned state.
Pink: the fluid state.
Pale green: the phase separated state.
Blue: the crystal state.
}
\label{fig:21}
\end{figure}

In Fig.~\ref{fig:21} we map out the different regimes in a dynamic
phase diagram as a function of ac frequency $\omega$ versus
ac amplitude $A$ for the system from Fig.~\ref{fig:20} in which both
species are subjected to ac driving that is out of phase.
We find a laned state, a fluid state, a phase separated state,
and a crystal state.
When the ac drive amplitude $A$ is small,
the repulsive particle-particle interaction dominates
the behavior and the system forms a rigid crystal
that rotates as a unit.
For large ac amplitude and low driving frequencies, a switching lanes state
emerges,
while for high driving frequencies the system forms a crystal due to the
fact that the circular orbits are too small to overlap with
each other.
The phase separated state appears between the crystal and fluid phases
as long as the ac driving amplitude is not too small.
When thermal fluctuations are added (not shown), we find a
temperature dependence similar to that described for the system with
dc driving on species B in Section V,
where the switching phase is able to persist 
for temperatures that are
smaller than the equilibrium melting temperature.

\section{Other Particle-Particle Interaction Potentials}

\begin{figure}
\includegraphics[width=\columnwidth]{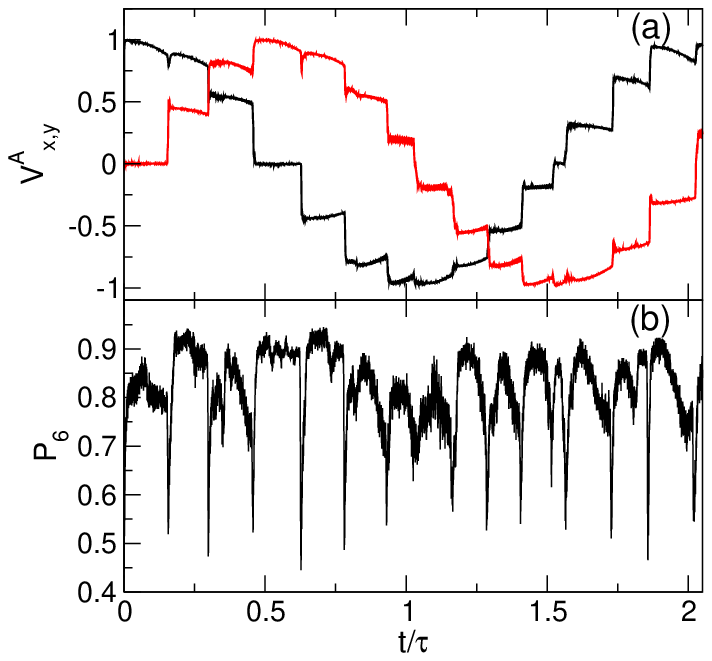}
\caption{(a) $V_x^A$ (black) and $V_x^B$ (red) vs $t/\tau$ for a system
where both species A and B are subjected to circular
driving that is out of phase by $\pi$ with amplitude $A=1.0$ and
frequency $\omega=1\times 10^{-6}$. Here the particle-particle interaction
potential has been changed to a repulsive Bessel function.
(b) The corresponding $P_6$ vs $t/\tau$.
}
\label{fig:22}
\end{figure}

\begin{figure}
\includegraphics[width=\columnwidth]{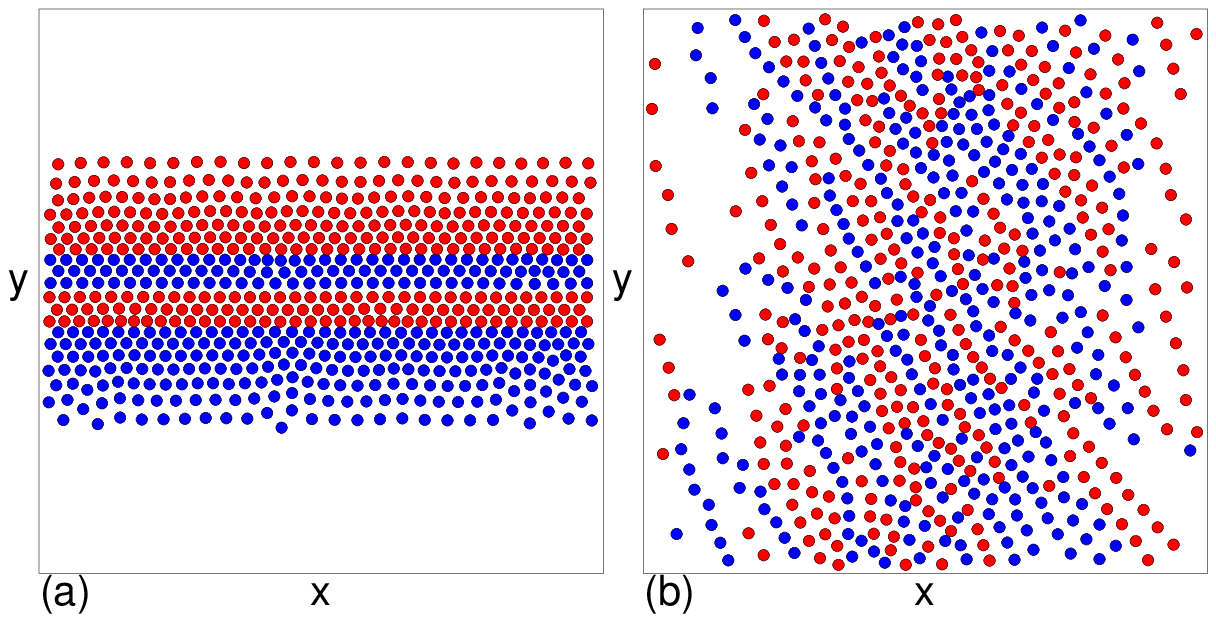}
\caption{Snapshots of particle positions for species A (blue) and B (red)
for the Bessel function system from Fig.~\ref{fig:22}
where both species A and B are subjected to circular driving that is out of
phase by $\pi$ with amplitude $A=1.0$ and frequency
$\omega=1\times 10^{-6}$.
(a) A dense laned state shortly before a switching event.
(b) A disordered state immediately after a switching event.
}
\label{fig:23}
\end{figure}

Up until this point we have described the behavior of systems with
Yukawa or screened Coulomb particle-particle interactions. To determine
whether the details of the interaction potential strongly affect the
behavior of the system, we replace the Yukawa potential with a
repulsive modified Bessel function $V(R_{ij})=F_0K_0(R_{ij})$,
where $F_0$ is a prefactor describing the strength of the interaction. This
potential describes the interactions between vortices in type-II
superconductors as well as stiff magnetic skyrmions. It decays exponentially
at larger distances.
With the Bessel function interaction
we find similar laning behavior as
for the particles with Yukawa interactions;
however, we must generally use a larger prefactor in the interaction
potential to find the same behavior in both systems at a fixed
particle density of $\rho=0.292.$
In Fig.~\ref{fig:22}(a,b) we plot
$V^{A}_{x,}$, $V^A_y$, and $P_6$ versus $t/\tau$
for a system in which both species A and B are subjected to out of phase
ac driving with amplitude
$A=1.0$ and frequency $\omega = 1\times10^{-6}$,
where the interaction potential is a Bessel function with
a prefactor of $F_{0} = 2.0$.
The jumps in the velocity, accompanied by dips in $P_6$, are very similar
to what we found for the same driving protocol with screened Yukawa
particles in Fig.~\ref{fig:18}.
In general, the lanes are more ordered in the Bessel function system than
in the Yukawa system, as indicated by the somewhat higher average value of
$P_6 \approx 0.9$ in Fig.~\ref{fig:22}(b) compared to Fig.~\ref{fig:18}(b).
In Fig.~\ref{fig:23}(a) we illustrate the particle positions in the
laned state during a strongly compressed portion of the drive cycle.
For the parameters used here, the Bessel function particle lattice is
softer than the Yukawa particle lattice from Fig.~\ref{fig:18},
so instead of forming uniformly spaced lanes, the particles assemble into
a wide but phase separated single lane state. The density gradient for the
Bessel function particles is weaker than that shown for the Yukawa system
due to the softness of the Bessel function particle lattice.
Figure~\ref{fig:23}(b) shows the particle configurations for the same
system immediately after a switching event where the particles are disordered.

\begin{figure}
\includegraphics[width=\columnwidth]{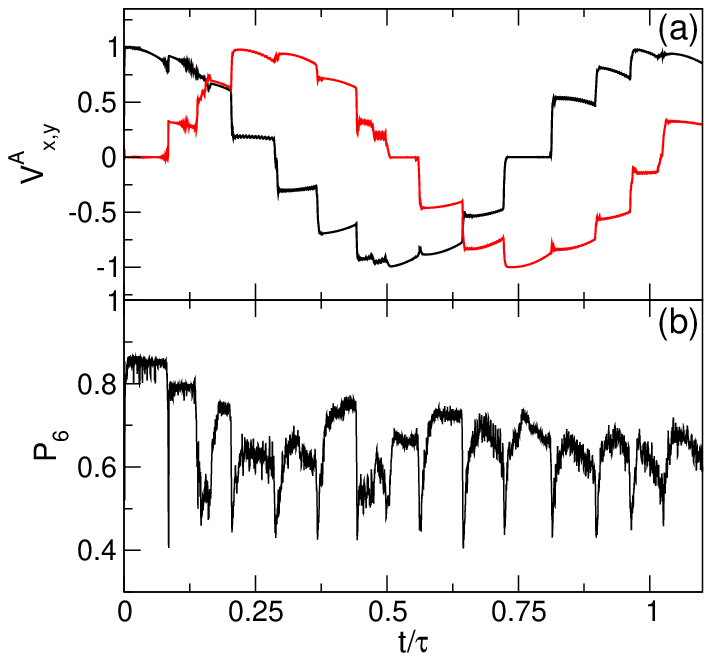}
\caption{(a) $V_x^A$ (black) and $V^A_y$ (red) vs $t/\tau$ for a system
with $\rho=0.2$ where both species A and B are subjected to circular
driving that is out of phase by $\pi$ with amplitude
$A=1.0$ and
frequency $\omega = 2\times10^{-6}$.
Here the particle-particle interaction potential has been changed to
long range Coulomb repulsion.
(b) The corresponding $P_6$ vs $t/\tau$.
}
\label{fig:24}
\end{figure}

The same dynamical behavior persists if we replace the Yukawa interaction with
a long range Coulomb interaction potential, $V(R_{ij})=R_{ij}$.
In Fig.~\ref{fig:24}(a) we plot $V^A_x$ and $V^A_y$ versus $t/\tau$ for
a Coulomb system in which species A and B are both driven by circular ac
drives that are out of phase by $\pi$ with amplitude $A=1.0$ and
frequency
$\omega = 2\times10^{-6}$.
Here we have lowered the particle density to $\rho = 0.2$ in
order to soften the lattice.
We once again observe the same type of switching events found
in the systems with shorter range repulsive interaction potentials.
The plot of the corresponding $P_6$ versus $t/\tau$ in
Fig.~\ref{fig:24}(b)
shows dips near the transition points.
For this set of parameters,
the Coulomb interacting particles are not as well
ordered along the
length of the lanes
compared to the shorter range potentials,
causing the overall values of $P_{6}$ to be lower,
but the same general behavior of switching among laning states accompanied
by disordering and ordering transitions still appears.
These results indicate
that the laning and switching behaviors
that occur for low ac driving frequencies
are robust features that can be observed
for many different kinds of particle-particle interaction potentials.

\section{Discussion and Connections to Stick-Slip Systems}

Our system can be viewed as exhibiting a form of stick-slip motion.
Typical studies of stick-slip motion 
\cite{BenDavid10,Tian16}
involve a driven system
with quenched disorder or friction
where, under an increasing drive, the system remains stuck until it breaks
away with a large velocity increase and then becomes stuck again.
In some cases, the system can
become fluid-like during the slip event.
In our system, the particles are always in motion, but as the ac drive
rotates into a new direction, the particles are not able to smoothly follow
the driving direction when lanes have formed along a direction that was
formerly aligned with the drive but is now no longer aligned with the
drive. The system can thus be regarded as stuck when the lanes are unable
to align with the drive.
As the misalignment between the direction of the rotating drive and the
orientation of the particle lanes increases,
this is akin to increasing the load on a stick-slip system,
and at some point the lanes break apart or slip,
leading to increased motion.
The system is disordered during this switching or slip event.
It then reorders to form a new laned state aligned with the new
driving direction,
which becomes a new stuck state as the drive continues to rotate, resulting
in another slip event. This process continues to repeat.
When thermal fluctuations are added to the system, the particles are always
sliding if the temperature is high enough, and the slip events
disappear.
In traditional stick-slip systems,
the frequency and size of the jumps
that occur often depend on the rate or speed of the loading.
In our system, this corresponds to the frequency of the ac drive, which
controls the rate at which the driving direction is varied.
When the frequency is high enough,
the system does not have time to reform a lane before the drive rotates
far enough into a new direction to destabilize the emerging lane,
which would correspond to a stick-slip system
not having time to return to
a stuck state in rapidly loaded stick-slip systems.

\section{Summary}

We have examined a binary system of particles with Yukawa interactions where
one or both of the particle species is subjected to a rotating ac
external driving force.
We first considered
systems where only species A is driven with a rotating drive while
species B is either passive or is subjected to a constant dc drive
applied in a fixed
direction.
At low ac drive frequencies,
the system forms a series of pattern-forming
or laned states.
When a lane forms, it aligns with the direction of the net drive at
that moment, and the direction of the lane then becomes
fixed or stuck to that particular direction.
As time progresses and the net direction of the drive rotates,
at some point the lanes destabilize and
break apart,
and the system form a new set of lanes aligned with the new direction of
the net drive.
This laning and switching process
repeats throughout the entire ac drive cycle
with up to twenty or more switching events occurring per cycle
depending on the parameters.
In a laned state, the system is partially ordered, while during a
switching or slipping event, the system is disordered.
Just before a switching event, the lanes become increasingly compressed,
while after the switching event, the density of the system is much more
uniform.
If the dc drive on species B is sufficiently large, a single lane
rather than a set of lanes appears, and it is oriented in the direction of
the species B dc driving force.
Lane switching events only occur for sufficiently low ac driving
frequencies where the lanes have time to form or reform
before the direction of the ac drive has changed significantly.
At higher ac driving frequencies,
the system forms a fluid state and there are no lanes or switching
among laned states.
At the highest ac driving frequencies,
when the radius of the circular orbit of the ac driven
species A particles is comparable to or smaller than
the average spacing between particles,
the system forms a phase-separated state or a crystal state.
We show that 
the switching phenomena and laning remain robust against the addition of
thermal fluctuations
up to the equilibrium melting temperature of the undriven lattice,
and can even persist to higher temperatures due to the local increase in
particle density during the portion of the ac driving cycle that compresses
the lane structure, which raises the effective melting temperature locally and
can cause local recrystallization.
We find similar laning phases
and switching behavior if both
particle species are subjected to the same ac driving force but are driven
out of phase with each other.
By varying the particle-particle interaction potential, we show that
the same switching dynamics
also occurs for particles with Bessel function
interactions and long range Coulomb interactions.
We explain that
the switching dynamics at low frequencies can be
viewed as an example of a stick-slip
phenomenon or a melting and recrystallization dynamics.
Since we have shown that our results remain robust
for a wide range of particle-particle interaction potentials,
it should be possible to realize the effects described here in a variety
of binary systems with repulsive interactions
where the two species respond differently to an external field or
other driving.

\begin{acknowledgements}
We gratefully acknowledge the support of the U.S. Department of
Energy through the LANL/LDRD program for this work.
This work was supported by the US Department of Energy through
the Los Alamos National Laboratory.  Los Alamos National Laboratory is
operated by Triad National Security, LLC, for the National Nuclear Security
Administration of the U. S. Department of Energy (Contract No. 892333218NCA000001).
\end{acknowledgements}

\bibliography{mybib}

\begin{thebibliography}{57}%
\makeatletter
\providecommand \@ifxundefined [1]{%
 \@ifx{#1\undefined}
}%
\providecommand \@ifnum [1]{%
 \ifnum #1\expandafter \@firstoftwo
 \else \expandafter \@secondoftwo
 \fi
}%
\providecommand \@ifx [1]{%
 \ifx #1\expandafter \@firstoftwo
 \else \expandafter \@secondoftwo
 \fi
}%
\providecommand \natexlab [1]{#1}%
\providecommand \enquote  [1]{``#1''}%
\providecommand \bibnamefont  [1]{#1}%
\providecommand \bibfnamefont [1]{#1}%
\providecommand \citenamefont [1]{#1}%
\providecommand \href@noop [0]{\@secondoftwo}%
\providecommand \href [0]{\begingroup \@sanitize@url \@href}%
\providecommand \@href[1]{\@@startlink{#1}\@@href}%
\providecommand \@@href[1]{\endgroup#1\@@endlink}%
\providecommand \@sanitize@url [0]{\catcode `\\12\catcode `\$12\catcode
  `\&12\catcode `\#12\catcode `\^12\catcode `\_12\catcode `\%12\relax}%
\providecommand \@@startlink[1]{}%
\providecommand \@@endlink[0]{}%
\providecommand \url  [0]{\begingroup\@sanitize@url \@url }%
\providecommand \@url [1]{\endgroup\@href {#1}{\urlprefix }}%
\providecommand \urlprefix  [0]{URL }%
\providecommand \Eprint [0]{\href }%
\providecommand \doibase [0]{https://doi.org/}%
\providecommand \selectlanguage [0]{\@gobble}%
\providecommand \bibinfo  [0]{\@secondoftwo}%
\providecommand \bibfield  [0]{\@secondoftwo}%
\providecommand \translation [1]{[#1]}%
\providecommand \BibitemOpen [0]{}%
\providecommand \bibitemStop [0]{}%
\providecommand \bibitemNoStop [0]{.\EOS\space}%
\providecommand \EOS [0]{\spacefactor3000\relax}%
\providecommand \BibitemShut  [1]{\csname bibitem#1\endcsname}%
\let\auto@bib@innerbib\@empty
\bibitem [{\citenamefont {Olson~Reichhardt}\ \emph {et~al.}(2010)\citenamefont
  {Olson~Reichhardt}, \citenamefont {Reichhardt},\ and\ \citenamefont
  {Bishop}}]{Reichhardt10}%
  \BibitemOpen
  \bibfield  {author} {\bibinfo {author} {\bibfnamefont {C.~J.}\ \bibnamefont
  {Olson~Reichhardt}}, \bibinfo {author} {\bibfnamefont {C.}~\bibnamefont
  {Reichhardt}},\ and\ \bibinfo {author} {\bibfnamefont {A.~R.}\ \bibnamefont
  {Bishop}},\ }\bibfield  {title} {\bibinfo {title} {Structural transitions,
  melting, and intermediate phases for stripe- and clump-forming systems},\
  }\href {https://doi.org/10.1103/PhysRevE.82.041502} {\bibfield  {journal}
  {\bibinfo  {journal} {Phys. Rev. E}\ }\textbf {\bibinfo {volume} {82}},\
  \bibinfo {pages} {041502} (\bibinfo {year} {2010})}\BibitemShut {NoStop}%
\bibitem [{\citenamefont {Meng}\ \emph {et~al.}(2017)\citenamefont {Meng},
  \citenamefont {Varney}, \citenamefont {Fangohr},\ and\ \citenamefont
  {Babaev}}]{Meng17}%
  \BibitemOpen
  \bibfield  {author} {\bibinfo {author} {\bibfnamefont {Q.}~\bibnamefont
  {Meng}}, \bibinfo {author} {\bibfnamefont {C.~N.}\ \bibnamefont {Varney}},
  \bibinfo {author} {\bibfnamefont {H.}~\bibnamefont {Fangohr}},\ and\ \bibinfo
  {author} {\bibfnamefont {E.}~\bibnamefont {Babaev}},\ }\bibfield  {title}
  {\bibinfo {title} {Phase diagrams of vortex matter with multi-scale
  inter-vortex interactions in layered superconductors},\ }\href
  {https://doi.org/10.1088/1361-648X/29/3/035602} {\bibfield  {journal}
  {\bibinfo  {journal} {J. Phys.: Condens. Matter}\ }\textbf {\bibinfo {volume}
  {29}},\ \bibinfo {pages} {035602} (\bibinfo {year} {2017})}\BibitemShut
  {NoStop}%
\bibitem [{\citenamefont {Xu}\ \emph {et~al.}(2021)\citenamefont {Xu},
  \citenamefont {Tang}, \citenamefont {Wang}, \citenamefont {Xu}, \citenamefont
  {Fang},\ and\ \citenamefont {Gu}}]{Xu21}%
  \BibitemOpen
  \bibfield  {author} {\bibinfo {author} {\bibfnamefont {X.~B.}\ \bibnamefont
  {Xu}}, \bibinfo {author} {\bibfnamefont {T.}~\bibnamefont {Tang}}, \bibinfo
  {author} {\bibfnamefont {Z.~H.}\ \bibnamefont {Wang}}, \bibinfo {author}
  {\bibfnamefont {X.~N.}\ \bibnamefont {Xu}}, \bibinfo {author} {\bibfnamefont
  {G.~Y.}\ \bibnamefont {Fang}},\ and\ \bibinfo {author} {\bibfnamefont
  {M.}~\bibnamefont {Gu}},\ }\bibfield  {title} {\bibinfo {title}
  {Nonequilibrium pattern formation in circularly confined two-dimensional
  systems with competing interactions},\ }\href
  {https://doi.org/10.1103/PhysRevE.103.012604} {\bibfield  {journal} {\bibinfo
   {journal} {Phys. Rev. E}\ }\textbf {\bibinfo {volume} {103}},\ \bibinfo
  {pages} {012604} (\bibinfo {year} {2021})}\BibitemShut {NoStop}%
\bibitem [{\citenamefont {Hooshanginejad}\ \emph {et~al.}(2024)\citenamefont
  {Hooshanginejad}, \citenamefont {Barotta}, \citenamefont {Spradlin},
  \citenamefont {Pucci}, \citenamefont {Hunt},\ and\ \citenamefont
  {Harris}}]{Hooshanginejad24}%
  \BibitemOpen
  \bibfield  {author} {\bibinfo {author} {\bibfnamefont {A.}~\bibnamefont
  {Hooshanginejad}}, \bibinfo {author} {\bibfnamefont {J.-W.}\ \bibnamefont
  {Barotta}}, \bibinfo {author} {\bibfnamefont {V.}~\bibnamefont {Spradlin}},
  \bibinfo {author} {\bibfnamefont {G.}~\bibnamefont {Pucci}}, \bibinfo
  {author} {\bibfnamefont {R.}~\bibnamefont {Hunt}},\ and\ \bibinfo {author}
  {\bibfnamefont {D.~M.}\ \bibnamefont {Harris}},\ }\bibfield  {title}
  {\bibinfo {title} {Interactions and pattern formation in a macroscopic
  magnetocapillary salr system of mermaid cereal},\ }\href
  {https://doi.org/10.1038/s41467-024-49754-4} {\bibfield  {journal} {\bibinfo
  {journal} {Nature Commun.}\ }\textbf {\bibinfo {volume} {15}},\ \bibinfo
  {pages} {5466} (\bibinfo {year} {2024})}\BibitemShut {NoStop}%
\bibitem [{\citenamefont {Seul}\ and\ \citenamefont {Wolfe}(1992)}]{Seul92}%
  \BibitemOpen
  \bibfield  {author} {\bibinfo {author} {\bibfnamefont {M.}~\bibnamefont
  {Seul}}\ and\ \bibinfo {author} {\bibfnamefont {R.}~\bibnamefont {Wolfe}},\
  }\bibfield  {title} {\bibinfo {title} {Evolution of disorder in magnetic
  stripe domains. {I. T}ransverse instabilities and disclination unbinding in
  lamellar patterns},\ }\href {https://doi.org/10.1103/PhysRevA.46.7519}
  {\bibfield  {journal} {\bibinfo  {journal} {Phys. Rev. A}\ }\textbf {\bibinfo
  {volume} {46}},\ \bibinfo {pages} {7519} (\bibinfo {year}
  {1992})}\BibitemShut {NoStop}%
\bibitem [{\citenamefont {Malescio}\ and\ \citenamefont
  {Pellicane}(2003)}]{Malescio03}%
  \BibitemOpen
  \bibfield  {author} {\bibinfo {author} {\bibfnamefont {G.}~\bibnamefont
  {Malescio}}\ and\ \bibinfo {author} {\bibfnamefont {G.}~\bibnamefont
  {Pellicane}},\ }\bibfield  {title} {\bibinfo {title} {Stripe phases from
  isotropic repulsive interactions},\ }\href {https://doi.org/10.1038/nmat820}
  {\bibfield  {journal} {\bibinfo  {journal} {Nature Mater.}\ }\textbf
  {\bibinfo {volume} {2}},\ \bibinfo {pages} {97} (\bibinfo {year}
  {2003})}\BibitemShut {NoStop}%
\bibitem [{\citenamefont {Massana-Cid}\ \emph {et~al.}(2021)\citenamefont
  {Massana-Cid}, \citenamefont {Levis}, \citenamefont {Josue~Hernandez},
  \citenamefont {Pagonabarraga},\ and\ \citenamefont {Tierno}}]{Massana21}%
  \BibitemOpen
  \bibfield  {author} {\bibinfo {author} {\bibfnamefont {H.}~\bibnamefont
  {Massana-Cid}}, \bibinfo {author} {\bibfnamefont {D.}~\bibnamefont {Levis}},
  \bibinfo {author} {\bibfnamefont {R.}~\bibnamefont {Josue~Hernandez}},
  \bibinfo {author} {\bibfnamefont {I.}~\bibnamefont {Pagonabarraga}},\ and\
  \bibinfo {author} {\bibfnamefont {P.~.}\ \bibnamefont {Tierno}},\ }\bibfield
  {title} {\bibinfo {title} {Arrested phase separation in chiral fluids of
  colloidal spinners},\ }\href
  {https://doi.org/10.1103/PhysRevResearch.3.L042021} {\bibfield  {journal}
  {\bibinfo  {journal} {Phys. Rev. Research}\ }\textbf {\bibinfo {volume}
  {3}},\ \bibinfo {pages} {L04201} (\bibinfo {year} {2021})}\BibitemShut
  {NoStop}%
\bibitem [{\citenamefont {Katzmeier}\ \emph {et~al.}(2022)\citenamefont
  {Katzmeier}, \citenamefont {Altaner}, \citenamefont {List}, \citenamefont
  {Gerland},\ and\ \citenamefont {Simmel}}]{Katzmeier22}%
  \BibitemOpen
  \bibfield  {author} {\bibinfo {author} {\bibfnamefont {F.}~\bibnamefont
  {Katzmeier}}, \bibinfo {author} {\bibfnamefont {B.}~\bibnamefont {Altaner}},
  \bibinfo {author} {\bibfnamefont {J.}~\bibnamefont {List}}, \bibinfo {author}
  {\bibfnamefont {U.}~\bibnamefont {Gerland}},\ and\ \bibinfo {author}
  {\bibfnamefont {F.~.}\ \bibnamefont {Simmel}},\ }\bibfield  {title} {\bibinfo
  {title} {Emergence of colloidal patterns in ac electric fields},\ }\href
  {https://doi.org/10.1103/PhysRevLett.128.058002} {\bibfield  {journal}
  {\bibinfo  {journal} {Phys. Rev. Lett.}\ }\textbf {\bibinfo {volume} {128}},\
  \bibinfo {pages} {058002} (\bibinfo {year} {2022})}\BibitemShut {NoStop}%
\bibitem [{\citenamefont {McDermott}\ \emph {et~al.}(2014)\citenamefont
  {McDermott}, \citenamefont {Reichhardt},\ and\ \citenamefont
  {Reichhardt}}]{McDermott14}%
  \BibitemOpen
  \bibfield  {author} {\bibinfo {author} {\bibfnamefont {D.}~\bibnamefont
  {McDermott}}, \bibinfo {author} {\bibfnamefont {C.~J.~O.}\ \bibnamefont
  {Reichhardt}},\ and\ \bibinfo {author} {\bibfnamefont {C.}~\bibnamefont
  {Reichhardt}},\ }\bibfield  {title} {\bibinfo {title} {Stripe systems with
  competing interactions on quasi-one dimensional periodic substrates},\ }\href
  {https://doi.org/10.1039/c4sm01341g} {\bibfield  {journal} {\bibinfo
  {journal} {Soft Matter}\ }\textbf {\bibinfo {volume} {10}},\ \bibinfo {pages}
  {6332} (\bibinfo {year} {2014})}\BibitemShut {NoStop}%
\bibitem [{\citenamefont {Reichhardt}\ and\ \citenamefont
  {Reichhardt}(2017)}]{Reichhardt17}%
  \BibitemOpen
  \bibfield  {author} {\bibinfo {author} {\bibfnamefont {C.}~\bibnamefont
  {Reichhardt}}\ and\ \bibinfo {author} {\bibfnamefont {C.~J.~O.}\ \bibnamefont
  {Reichhardt}},\ }\bibfield  {title} {\bibinfo {title} {Depinning and
  nonequilibrium dynamic phases of particle assemblies driven over random and
  ordered substrates: a review},\ }\href
  {https://doi.org/10.1088/1361-6633/80/2/026501} {\bibfield  {journal}
  {\bibinfo  {journal} {Rep. Prog. Phys.}\ }\textbf {\bibinfo {volume} {80}},\
  \bibinfo {pages} {026501} (\bibinfo {year} {2017})}\BibitemShut {NoStop}%
\bibitem [{\citenamefont {Dzubiella}\ \emph {et~al.}(2002)\citenamefont
  {Dzubiella}, \citenamefont {Hoffmann},\ and\ \citenamefont
  {L\"owen}}]{Dzubiella02a}%
  \BibitemOpen
  \bibfield  {author} {\bibinfo {author} {\bibfnamefont {J.}~\bibnamefont
  {Dzubiella}}, \bibinfo {author} {\bibfnamefont {G.~P.}\ \bibnamefont
  {Hoffmann}},\ and\ \bibinfo {author} {\bibfnamefont {H.}~\bibnamefont
  {L\"owen}},\ }\bibfield  {title} {\bibinfo {title} {Lane formation in
  colloidal mixtures driven by an external field},\ }\href
  {https://doi.org/10.1103/PhysRevE.65.021402} {\bibfield  {journal} {\bibinfo
  {journal} {Phys. Rev. E}\ }\textbf {\bibinfo {volume} {65}},\ \bibinfo
  {pages} {021402} (\bibinfo {year} {2002})}\BibitemShut {NoStop}%
\bibitem [{\citenamefont {Rex}\ and\ \citenamefont {L\"owen}(2007)}]{Rex07}%
  \BibitemOpen
  \bibfield  {author} {\bibinfo {author} {\bibfnamefont {M.}~\bibnamefont
  {Rex}}\ and\ \bibinfo {author} {\bibfnamefont {H.}~\bibnamefont {L\"owen}},\
  }\bibfield  {title} {\bibinfo {title} {Lane formation in oppositely charged
  colloids driven by an electric field: Chaining and two-dimensional
  crystallization},\ }\href {https://doi.org/10.1103/PhysRevE.75.051402}
  {\bibfield  {journal} {\bibinfo  {journal} {Phys. Rev. E}\ }\textbf {\bibinfo
  {volume} {75}},\ \bibinfo {pages} {051402} (\bibinfo {year}
  {2007})}\BibitemShut {NoStop}%
\bibitem [{\citenamefont {Glanz}\ and\ \citenamefont {L{\"
  o}wen}(2012)}]{Glanz12}%
  \BibitemOpen
  \bibfield  {author} {\bibinfo {author} {\bibfnamefont {T.}~\bibnamefont
  {Glanz}}\ and\ \bibinfo {author} {\bibfnamefont {H.}~\bibnamefont {L{\"
  o}wen}},\ }\bibfield  {title} {\bibinfo {title} {The nature of the laning
  transition in two dimensions},\ }\href
  {https://doi.org/10.1088/0953-8984/24/46/464114} {\bibfield  {journal}
  {\bibinfo  {journal} {J. Phys: Condens. Matter}\ }\textbf {\bibinfo {volume}
  {24}},\ \bibinfo {pages} {464114} (\bibinfo {year} {2012})}\BibitemShut
  {NoStop}%
\bibitem [{\citenamefont {Wysocki}\ and\ \citenamefont
  {L\"owen}(2009)}]{Wysocki09}%
  \BibitemOpen
  \bibfield  {author} {\bibinfo {author} {\bibfnamefont {A.}~\bibnamefont
  {Wysocki}}\ and\ \bibinfo {author} {\bibfnamefont {H.}~\bibnamefont
  {L\"owen}},\ }\bibfield  {title} {\bibinfo {title} {Oscillatory driven
  colloidal binary mixtures: Axial segregation versus laning},\ }\href
  {https://doi.org/10.1103/PhysRevE.79.041408} {\bibfield  {journal} {\bibinfo
  {journal} {Phys. Rev. E}\ }\textbf {\bibinfo {volume} {79}},\ \bibinfo
  {pages} {041408} (\bibinfo {year} {2009})}\BibitemShut {NoStop}%
\bibitem [{\citenamefont {Vissers}\ \emph {et~al.}(2011)\citenamefont
  {Vissers}, \citenamefont {van Blaaderen},\ and\ \citenamefont
  {Imhof}}]{Vissers11a}%
  \BibitemOpen
  \bibfield  {author} {\bibinfo {author} {\bibfnamefont {T.}~\bibnamefont
  {Vissers}}, \bibinfo {author} {\bibfnamefont {A.}~\bibnamefont {van
  Blaaderen}},\ and\ \bibinfo {author} {\bibfnamefont {A.}~\bibnamefont
  {Imhof}},\ }\bibfield  {title} {\bibinfo {title} {Band formation in mixtures
  of oppositely charged colloids driven by an ac electric field},\ }\href
  {https://doi.org/10.1103/PhysRevLett.106.228303} {\bibfield  {journal}
  {\bibinfo  {journal} {Phys. Rev. Lett.}\ }\textbf {\bibinfo {volume} {106}},\
  \bibinfo {pages} {228303} (\bibinfo {year} {2011})}\BibitemShut {NoStop}%
\bibitem [{\citenamefont {Li}\ \emph {et~al.}(2021)\citenamefont {Li},
  \citenamefont {Wang}, \citenamefont {Shi}, \citenamefont {Gao}, \citenamefont
  {Shi}, \citenamefont {Woodward},\ and\ \citenamefont {Forsman}}]{Li21}%
  \BibitemOpen
  \bibfield  {author} {\bibinfo {author} {\bibfnamefont {B.}~\bibnamefont
  {Li}}, \bibinfo {author} {\bibfnamefont {Y.-L.}\ \bibnamefont {Wang}},
  \bibinfo {author} {\bibfnamefont {G.}~\bibnamefont {Shi}}, \bibinfo {author}
  {\bibfnamefont {Y.}~\bibnamefont {Gao}}, \bibinfo {author} {\bibfnamefont
  {X.}~\bibnamefont {Shi}}, \bibinfo {author} {\bibfnamefont {C.~E.}\
  \bibnamefont {Woodward}},\ and\ \bibinfo {author} {\bibfnamefont
  {J.}~\bibnamefont {Forsman}},\ }\bibfield  {title} {\bibinfo {title} {Phase
  transitions of oppositely charged colloidal particles driven by alternating
  current electric field},\ }\href {https://doi.org/10.1021/acsnano.0c04095}
  {\bibfield  {journal} {\bibinfo  {journal} {ACS Nano}\ }\textbf {\bibinfo
  {volume} {15}},\ \bibinfo {pages} {2363} (\bibinfo {year}
  {2021})}\BibitemShut {NoStop}%
\bibitem [{\citenamefont {Tierno}\ \emph {et~al.}(2007)\citenamefont {Tierno},
  \citenamefont {Muruganathan},\ and\ \citenamefont {Fischer}}]{Tierno07}%
  \BibitemOpen
  \bibfield  {author} {\bibinfo {author} {\bibfnamefont {P.}~\bibnamefont
  {Tierno}}, \bibinfo {author} {\bibfnamefont {R.}~\bibnamefont
  {Muruganathan}},\ and\ \bibinfo {author} {\bibfnamefont {T.~M.}\ \bibnamefont
  {Fischer}},\ }\bibfield  {title} {\bibinfo {title} {Viscoelasticity of
  dynamically self-assembled paramagnetic colloidal clusters},\ }\href
  {https://doi.org/10.1103/PhysRevLett.98.028301} {\bibfield  {journal}
  {\bibinfo  {journal} {Phys. Rev. Lett.}\ }\textbf {\bibinfo {volume} {98}},\
  \bibinfo {pages} {028301} (\bibinfo {year} {2007})}\BibitemShut {NoStop}%
\bibitem [{\citenamefont {van Zuiden}\ \emph {et~al.}(2016)\citenamefont {van
  Zuiden}, \citenamefont {Paulose}, \citenamefont {Irvine}, \citenamefont
  {Bartolo},\ and\ \citenamefont {Vitelli}}]{vanZuiden16}%
  \BibitemOpen
  \bibfield  {author} {\bibinfo {author} {\bibfnamefont {B.~C.}\ \bibnamefont
  {van Zuiden}}, \bibinfo {author} {\bibfnamefont {J.}~\bibnamefont {Paulose}},
  \bibinfo {author} {\bibfnamefont {W.~T.~M.}\ \bibnamefont {Irvine}}, \bibinfo
  {author} {\bibfnamefont {D.}~\bibnamefont {Bartolo}},\ and\ \bibinfo {author}
  {\bibfnamefont {V.}~\bibnamefont {Vitelli}},\ }\bibfield  {title} {\bibinfo
  {title} {Spatiotemporal order and emergent edge currents in active spinner
  materials},\ }\href {https://doi.org/10.1073/pnas.1609572113} {\bibfield
  {journal} {\bibinfo  {journal} {Proc. Natl. Acad. Sci. (USA)}\ }\textbf
  {\bibinfo {volume} {113}},\ \bibinfo {pages} {12919} (\bibinfo {year}
  {2016})}\BibitemShut {NoStop}%
\bibitem [{\citenamefont {Han}\ \emph {et~al.}(2017)\citenamefont {Han},
  \citenamefont {Yan}, \citenamefont {Granick},\ and\ \citenamefont
  {Luijten}}]{Han17}%
  \BibitemOpen
  \bibfield  {author} {\bibinfo {author} {\bibfnamefont {M.}~\bibnamefont
  {Han}}, \bibinfo {author} {\bibfnamefont {J.}~\bibnamefont {Yan}}, \bibinfo
  {author} {\bibfnamefont {S.}~\bibnamefont {Granick}},\ and\ \bibinfo {author}
  {\bibfnamefont {E.}~\bibnamefont {Luijten}},\ }\bibfield  {title} {\bibinfo
  {title} {Effective temperature concept evaluated in an active colloid
  mixture},\ }\href {https://doi.org/10.1073/pnas.1706702114} {\bibfield
  {journal} {\bibinfo  {journal} {Proc. Natl. Acad. Sci. (USA)}\ }\textbf
  {\bibinfo {volume} {114}},\ \bibinfo {pages} {7513} (\bibinfo {year}
  {2017})}\BibitemShut {NoStop}%
\bibitem [{\citenamefont {Reichhardt}\ and\ \citenamefont
  {Reichhardt}(2019{\natexlab{a}})}]{Reichhardt19}%
  \BibitemOpen
  \bibfield  {author} {\bibinfo {author} {\bibfnamefont {C.}~\bibnamefont
  {Reichhardt}}\ and\ \bibinfo {author} {\bibfnamefont {C.~J.~O.}\ \bibnamefont
  {Reichhardt}},\ }\bibfield  {title} {\bibinfo {title} {Reversibility, pattern
  formation, and edge transport in active chiral and passive disk mixtures},\
  }\href {https://doi.org/10.1063/1.5085209} {\bibfield  {journal} {\bibinfo
  {journal} {J. Chem. Phys.}\ }\textbf {\bibinfo {volume} {150}},\ \bibinfo
  {pages} {064905} (\bibinfo {year} {2019}{\natexlab{a}})}\BibitemShut
  {NoStop}%
\bibitem [{\citenamefont {Reichhardt}\ and\ \citenamefont
  {Reichhardt}(2019{\natexlab{b}})}]{Reichhardt19b}%
  \BibitemOpen
  \bibfield  {author} {\bibinfo {author} {\bibfnamefont {C.~J.~O.}\
  \bibnamefont {Reichhardt}}\ and\ \bibinfo {author} {\bibfnamefont
  {C.}~\bibnamefont {Reichhardt}},\ }\bibfield  {title} {\bibinfo {title}
  {Disordering, clustering, and laning transitions in particle systems with
  dispersion in the {M}agnus term},\ }\href
  {https://doi.org/10.1103/PhysRevE.99.012606} {\bibfield  {journal} {\bibinfo
  {journal} {Phys. Rev. E}\ }\textbf {\bibinfo {volume} {99}},\ \bibinfo
  {pages} {012606} (\bibinfo {year} {2019}{\natexlab{b}})}\BibitemShut
  {NoStop}%
\bibitem [{\citenamefont {Chakrabarti}\ \emph {et~al.}(2004)\citenamefont
  {Chakrabarti}, \citenamefont {Dzubiella},\ and\ \citenamefont
  {L\"owen}}]{Chakrabarti04}%
  \BibitemOpen
  \bibfield  {author} {\bibinfo {author} {\bibfnamefont {J.}~\bibnamefont
  {Chakrabarti}}, \bibinfo {author} {\bibfnamefont {J.}~\bibnamefont
  {Dzubiella}},\ and\ \bibinfo {author} {\bibfnamefont {H.}~\bibnamefont
  {L\"owen}},\ }\bibfield  {title} {\bibinfo {title} {Reentrance effect in the
  lane formation of driven colloids},\ }\href
  {https://doi.org/10.1103/PhysRevE.70.012401} {\bibfield  {journal} {\bibinfo
  {journal} {Phys. Rev. E}\ }\textbf {\bibinfo {volume} {70}},\ \bibinfo
  {pages} {012401} (\bibinfo {year} {2004})}\BibitemShut {NoStop}%
\bibitem [{\citenamefont {Ikeda}\ \emph {et~al.}(2012)\citenamefont {Ikeda},
  \citenamefont {Wada},\ and\ \citenamefont {Hayakawa}}]{Ikeda12}%
  \BibitemOpen
  \bibfield  {author} {\bibinfo {author} {\bibfnamefont {M.}~\bibnamefont
  {Ikeda}}, \bibinfo {author} {\bibfnamefont {H.}~\bibnamefont {Wada}},\ and\
  \bibinfo {author} {\bibfnamefont {H.}~\bibnamefont {Hayakawa}},\ }\bibfield
  {title} {\bibinfo {title} {Instabilities and turbulence-like dynamics in an
  oppositely driven binary particle mixture},\ }\href
  {https://doi.org/10.1209/0295-5075/99/68005} {\bibfield  {journal} {\bibinfo
  {journal} {EPL}\ }\textbf {\bibinfo {volume} {99}},\ \bibinfo {pages} {68005}
  (\bibinfo {year} {2012})}\BibitemShut {NoStop}%
\bibitem [{\citenamefont {Klymko}\ \emph {et~al.}(2016)\citenamefont {Klymko},
  \citenamefont {Geissler},\ and\ \citenamefont {Whitelam}}]{Klymko16}%
  \BibitemOpen
  \bibfield  {author} {\bibinfo {author} {\bibfnamefont {K.}~\bibnamefont
  {Klymko}}, \bibinfo {author} {\bibfnamefont {P.~L.}\ \bibnamefont
  {Geissler}},\ and\ \bibinfo {author} {\bibfnamefont {S.}~\bibnamefont
  {Whitelam}},\ }\bibfield  {title} {\bibinfo {title} {Microscopic origin and
  macroscopic implications of lane formation in mixtures of oppositely driven
  particles},\ }\href {https://doi.org/10.1103/PhysRevE.94.022608} {\bibfield
  {journal} {\bibinfo  {journal} {Phys. Rev. E}\ }\textbf {\bibinfo {volume}
  {94}},\ \bibinfo {pages} {022608} (\bibinfo {year} {2016})}\BibitemShut
  {NoStop}%
\bibitem [{\citenamefont {Poncet}\ \emph {et~al.}(2017)\citenamefont {Poncet},
  \citenamefont {B\'enichou}, \citenamefont {D\'emery},\ and\ \citenamefont
  {Oshanin}}]{Poncet17}%
  \BibitemOpen
  \bibfield  {author} {\bibinfo {author} {\bibfnamefont {A.}~\bibnamefont
  {Poncet}}, \bibinfo {author} {\bibfnamefont {O.}~\bibnamefont {B\'enichou}},
  \bibinfo {author} {\bibfnamefont {V.}~\bibnamefont {D\'emery}},\ and\
  \bibinfo {author} {\bibfnamefont {G.}~\bibnamefont {Oshanin}},\ }\bibfield
  {title} {\bibinfo {title} {Universal long ranged correlations in driven
  binary mixtures},\ }\href {https://doi.org/10.1103/PhysRevLett.118.118002}
  {\bibfield  {journal} {\bibinfo  {journal} {Phys. Rev. Lett.}\ }\textbf
  {\bibinfo {volume} {118}},\ \bibinfo {pages} {118002} (\bibinfo {year}
  {2017})}\BibitemShut {NoStop}%
\bibitem [{\citenamefont {Glanz}\ \emph {et~al.}(2016)\citenamefont {Glanz},
  \citenamefont {Wittkowski},\ and\ \citenamefont {L\"owen}}]{Glanz16}%
  \BibitemOpen
  \bibfield  {author} {\bibinfo {author} {\bibfnamefont {T.}~\bibnamefont
  {Glanz}}, \bibinfo {author} {\bibfnamefont {R.}~\bibnamefont {Wittkowski}},\
  and\ \bibinfo {author} {\bibfnamefont {H.}~\bibnamefont {L\"owen}},\
  }\bibfield  {title} {\bibinfo {title} {Symmetry breaking in clogging for
  oppositely driven particles},\ }\href
  {https://doi.org/10.1103/PhysRevE.94.052606} {\bibfield  {journal} {\bibinfo
  {journal} {Phys. Rev. E}\ }\textbf {\bibinfo {volume} {94}},\ \bibinfo
  {pages} {052606} (\bibinfo {year} {2016})}\BibitemShut {NoStop}%
\bibitem [{\citenamefont {Reichhardt}\ and\ \citenamefont
  {Reichhardt}(2018)}]{Reichhardt18}%
  \BibitemOpen
  \bibfield  {author} {\bibinfo {author} {\bibfnamefont {C.}~\bibnamefont
  {Reichhardt}}\ and\ \bibinfo {author} {\bibfnamefont {C.~J.~O.}\ \bibnamefont
  {Reichhardt}},\ }\bibfield  {title} {\bibinfo {title} {Velocity force curves,
  laning, and jamming for oppositely driven disk systems},\ }\href
  {https://doi.org/10.1039/c7sm02162c} {\bibfield  {journal} {\bibinfo
  {journal} {Soft Matter}\ }\textbf {\bibinfo {volume} {14}},\ \bibinfo {pages}
  {490} (\bibinfo {year} {2018})}\BibitemShut {NoStop}%
\bibitem [{\citenamefont {Helbing}\ \emph {et~al.}(2000)\citenamefont
  {Helbing}, \citenamefont {Farkas},\ and\ \citenamefont {Vicsek}}]{Helbing00}%
  \BibitemOpen
  \bibfield  {author} {\bibinfo {author} {\bibfnamefont {D.}~\bibnamefont
  {Helbing}}, \bibinfo {author} {\bibfnamefont {I.~J.}\ \bibnamefont
  {Farkas}},\ and\ \bibinfo {author} {\bibfnamefont {T.}~\bibnamefont
  {Vicsek}},\ }\bibfield  {title} {\bibinfo {title} {Freezing by heating in a
  driven mesoscopic system},\ }\href
  {https://doi.org/10.1103/PhysRevLett.84.1240} {\bibfield  {journal} {\bibinfo
   {journal} {Phys. Rev. Lett.}\ }\textbf {\bibinfo {volume} {84}},\ \bibinfo
  {pages} {1240} (\bibinfo {year} {2000})}\BibitemShut {NoStop}%
\bibitem [{\citenamefont {Bain}\ and\ \citenamefont {Bartolo}(2017)}]{Bain17}%
  \BibitemOpen
  \bibfield  {author} {\bibinfo {author} {\bibfnamefont {N.}~\bibnamefont
  {Bain}}\ and\ \bibinfo {author} {\bibfnamefont {D.}~\bibnamefont {Bartolo}},\
  }\bibfield  {title} {\bibinfo {title} {Critical mingling and universal
  correlations in model binary active liquids},\ }\href
  {https://doi.org/10.1038/ncomms15969} {\bibfield  {journal} {\bibinfo
  {journal} {Nature Commun.}\ }\textbf {\bibinfo {volume} {8}},\ \bibinfo
  {pages} {15969} (\bibinfo {year} {2017})}\BibitemShut {NoStop}%
\bibitem [{\citenamefont {Yu}\ \emph {et~al.}(2022)\citenamefont {Yu},
  \citenamefont {Thijssen},\ and\ \citenamefont {Jack}}]{Yu22}%
  \BibitemOpen
  \bibfield  {author} {\bibinfo {author} {\bibfnamefont {H.}~\bibnamefont
  {Yu}}, \bibinfo {author} {\bibfnamefont {K.}~\bibnamefont {Thijssen}},\ and\
  \bibinfo {author} {\bibfnamefont {R.~L.}\ \bibnamefont {Jack}},\ }\bibfield
  {title} {\bibinfo {title} {Perpendicular and parallel phase separation in
  two-species driven diffusive lattice gases},\ }\href
  {https://doi.org/10.1103/PhysRevE.106.024129} {\bibfield  {journal} {\bibinfo
   {journal} {Phys. Rev. E}\ }\textbf {\bibinfo {volume} {106}},\ \bibinfo
  {pages} {024129} (\bibinfo {year} {2022})}\BibitemShut {NoStop}%
\bibitem [{\citenamefont {Yu}\ and\ \citenamefont {Jack}(2024)}]{Yu24}%
  \BibitemOpen
  \bibfield  {author} {\bibinfo {author} {\bibfnamefont {H.}~\bibnamefont
  {Yu}}\ and\ \bibinfo {author} {\bibfnamefont {R.~L.}\ \bibnamefont {Jack}},\
  }\bibfield  {title} {\bibinfo {title} {Competition between lanes and
  transient jammed clusters in driven binary mixtures},\ }\href
  {https://doi.org/10.1103/PhysRevE.109.024123} {\bibfield  {journal} {\bibinfo
   {journal} {Phys. Rev. E}\ }\textbf {\bibinfo {volume} {109}},\ \bibinfo
  {pages} {024123} (\bibinfo {year} {2024})}\BibitemShut {NoStop}%
\bibitem [{\citenamefont {Reichhardt}\ and\ \citenamefont
  {Reichhardt}(2006)}]{Reichhardt06}%
  \BibitemOpen
  \bibfield  {author} {\bibinfo {author} {\bibfnamefont {C.}~\bibnamefont
  {Reichhardt}}\ and\ \bibinfo {author} {\bibfnamefont {C.~J.~O.}\ \bibnamefont
  {Reichhardt}},\ }\bibfield  {title} {\bibinfo {title} {Cooperative behavior
  and pattern formation in mixtures of driven and nondriven colloidal
  assemblies},\ }\href {https://doi.org/10.1103/PhysRevE.74.011403} {\bibfield
  {journal} {\bibinfo  {journal} {Phys. Rev. E}\ }\textbf {\bibinfo {volume}
  {74}},\ \bibinfo {pages} {011403} (\bibinfo {year} {2006})}\BibitemShut
  {NoStop}%
\bibitem [{\citenamefont {Kogler}\ and\ \citenamefont
  {Klapp}(2015)}]{Kogler15}%
  \BibitemOpen
  \bibfield  {author} {\bibinfo {author} {\bibfnamefont {F.}~\bibnamefont
  {Kogler}}\ and\ \bibinfo {author} {\bibfnamefont {S.~H.~L.}\ \bibnamefont
  {Klapp}},\ }\bibfield  {title} {\bibinfo {title} {Lane formation in a system
  of dipolar microswimmers},\ }\href
  {https://doi.org/10.1209/0295-5075/110/10004} {\bibfield  {journal} {\bibinfo
   {journal} {EPL}\ }\textbf {\bibinfo {volume} {110}},\ \bibinfo {pages}
  {10004} (\bibinfo {year} {2015})}\BibitemShut {NoStop}%
\bibitem [{\citenamefont {W\"achtler}\ \emph {et~al.}(2016)\citenamefont
  {W\"achtler}, \citenamefont {Kogler},\ and\ \citenamefont
  {Klapp}}]{Wachtler16}%
  \BibitemOpen
  \bibfield  {author} {\bibinfo {author} {\bibfnamefont {C.~W.}\ \bibnamefont
  {W\"achtler}}, \bibinfo {author} {\bibfnamefont {F.}~\bibnamefont {Kogler}},\
  and\ \bibinfo {author} {\bibfnamefont {S.~H.~L.}\ \bibnamefont {Klapp}},\
  }\bibfield  {title} {\bibinfo {title} {Lane formation in a driven attractive
  fluid},\ }\href {https://doi.org/10.1103/PhysRevE.94.052603} {\bibfield
  {journal} {\bibinfo  {journal} {Phys. Rev. E}\ }\textbf {\bibinfo {volume}
  {94}},\ \bibinfo {pages} {052603} (\bibinfo {year} {2016})}\BibitemShut
  {NoStop}%
\bibitem [{\citenamefont {Reichhardt}\ and\ \citenamefont
  {Reichhardt}(2026)}]{Reichhardt26}%
  \BibitemOpen
  \bibfield  {author} {\bibinfo {author} {\bibfnamefont {C.}~\bibnamefont
  {Reichhardt}}\ and\ \bibinfo {author} {\bibfnamefont {C.~J.~O.}\ \bibnamefont
  {Reichhardt}},\ }\bibfield  {title} {\bibinfo {title} {Hysteresis, laning,
  and negative drag in binary systems with opposite and perpendicular
  driving},\ }\Eprint {https://arxiv.org/abs/2512.13925} {arXiv:2512.13925}
  (\bibinfo {year} {2026})\BibitemShut {NoStop}%
\bibitem [{\citenamefont {Leunissen}\ \emph {et~al.}(2005)\citenamefont
  {Leunissen}, \citenamefont {Christova}, \citenamefont {Hynninen},
  \citenamefont {Royall}, \citenamefont {Campbell}, \citenamefont {Imhof},
  \citenamefont {Dijkstra}, \citenamefont {van Roij},\ and\ \citenamefont {van
  Blaaderen}}]{Leunissen05}%
  \BibitemOpen
  \bibfield  {author} {\bibinfo {author} {\bibfnamefont {M.~E.}\ \bibnamefont
  {Leunissen}}, \bibinfo {author} {\bibfnamefont {C.~G.}\ \bibnamefont
  {Christova}}, \bibinfo {author} {\bibfnamefont {A.~P.}\ \bibnamefont
  {Hynninen}}, \bibinfo {author} {\bibfnamefont {C.~P.}\ \bibnamefont
  {Royall}}, \bibinfo {author} {\bibfnamefont {A.~I.}\ \bibnamefont
  {Campbell}}, \bibinfo {author} {\bibfnamefont {A.}~\bibnamefont {Imhof}},
  \bibinfo {author} {\bibfnamefont {M.}~\bibnamefont {Dijkstra}}, \bibinfo
  {author} {\bibfnamefont {R.}~\bibnamefont {van Roij}},\ and\ \bibinfo
  {author} {\bibfnamefont {A.}~\bibnamefont {van Blaaderen}},\ }\bibfield
  {title} {\bibinfo {title} {Ionic colloidal crystals of oppositely charged
  particles},\ }\href {https://doi.org/10.1038/nature03946} {\bibfield
  {journal} {\bibinfo  {journal} {Nature (London)}\ }\textbf {\bibinfo {volume}
  {437}},\ \bibinfo {pages} {235} (\bibinfo {year} {2005})}\BibitemShut
  {NoStop}%
\bibitem [{\citenamefont {S\"utterlin}\ \emph {et~al.}(2009)\citenamefont
  {S\"utterlin}, \citenamefont {Wysocki}, \citenamefont {Ivlev}, \citenamefont
  {R\"ath}, \citenamefont {Thomas}, \citenamefont {Rubin-Zuzic}, \citenamefont
  {Goedheer}, \citenamefont {Fortov}, \citenamefont {Lipaev}, \citenamefont
  {Molotkov}, \citenamefont {Petrov}, \citenamefont {Morfill},\ and\
  \citenamefont {L\"owen}}]{Sutterlin09}%
  \BibitemOpen
  \bibfield  {author} {\bibinfo {author} {\bibfnamefont {K.~R.}\ \bibnamefont
  {S\"utterlin}}, \bibinfo {author} {\bibfnamefont {A.}~\bibnamefont
  {Wysocki}}, \bibinfo {author} {\bibfnamefont {A.~V.}\ \bibnamefont {Ivlev}},
  \bibinfo {author} {\bibfnamefont {C.}~\bibnamefont {R\"ath}}, \bibinfo
  {author} {\bibfnamefont {H.~M.}\ \bibnamefont {Thomas}}, \bibinfo {author}
  {\bibfnamefont {M.}~\bibnamefont {Rubin-Zuzic}}, \bibinfo {author}
  {\bibfnamefont {W.~J.}\ \bibnamefont {Goedheer}}, \bibinfo {author}
  {\bibfnamefont {V.~E.}\ \bibnamefont {Fortov}}, \bibinfo {author}
  {\bibfnamefont {A.~M.}\ \bibnamefont {Lipaev}}, \bibinfo {author}
  {\bibfnamefont {V.~I.}\ \bibnamefont {Molotkov}}, \bibinfo {author}
  {\bibfnamefont {O.~F.}\ \bibnamefont {Petrov}}, \bibinfo {author}
  {\bibfnamefont {G.~E.}\ \bibnamefont {Morfill}},\ and\ \bibinfo {author}
  {\bibfnamefont {H.}~\bibnamefont {L\"owen}},\ }\bibfield  {title} {\bibinfo
  {title} {Dynamics of lane formation in driven binary complex plasmas},\
  }\href {https://doi.org/10.1103/PhysRevLett.102.085003} {\bibfield  {journal}
  {\bibinfo  {journal} {Phys. Rev. Lett.}\ }\textbf {\bibinfo {volume} {102}},\
  \bibinfo {pages} {085003} (\bibinfo {year} {2009})}\BibitemShut {NoStop}%
\bibitem [{\citenamefont {Isele}\ \emph {et~al.}(2023)\citenamefont {Isele},
  \citenamefont {Hofmann}, \citenamefont {Erbe}, \citenamefont {Leiderer},\
  and\ \citenamefont {Nielaba}}]{Isele23}%
  \BibitemOpen
  \bibfield  {author} {\bibinfo {author} {\bibfnamefont {M.}~\bibnamefont
  {Isele}}, \bibinfo {author} {\bibfnamefont {K.}~\bibnamefont {Hofmann}},
  \bibinfo {author} {\bibfnamefont {A.}~\bibnamefont {Erbe}}, \bibinfo {author}
  {\bibfnamefont {P.}~\bibnamefont {Leiderer}},\ and\ \bibinfo {author}
  {\bibfnamefont {P.}~\bibnamefont {Nielaba}},\ }\bibfield  {title} {\bibinfo
  {title} {Lane formation of colloidal particles driven in parallel by
  gravity},\ }\href {https://doi.org/10.1103/PhysRevE.108.034607} {\bibfield
  {journal} {\bibinfo  {journal} {Phys. Rev. E}\ }\textbf {\bibinfo {volume}
  {108}},\ \bibinfo {pages} {034607} (\bibinfo {year} {2023})}\BibitemShut
  {NoStop}%
\bibitem [{\citenamefont {Reichhardt}\ \emph {et~al.}(2018)\citenamefont
  {Reichhardt}, \citenamefont {Thibault}, \citenamefont {Papanikolaou},\ and\
  \citenamefont {Reichhardt}}]{Reichhardt18b}%
  \BibitemOpen
  \bibfield  {author} {\bibinfo {author} {\bibfnamefont {C.}~\bibnamefont
  {Reichhardt}}, \bibinfo {author} {\bibfnamefont {J.}~\bibnamefont
  {Thibault}}, \bibinfo {author} {\bibfnamefont {S.}~\bibnamefont
  {Papanikolaou}},\ and\ \bibinfo {author} {\bibfnamefont {C.~J.~O.}\
  \bibnamefont {Reichhardt}},\ }\bibfield  {title} {\bibinfo {title} {Laning
  and clustering transitions in driven binary active matter systems},\ }\href
  {https://doi.org/10.1103/PhysRevE.98.022603} {\bibfield  {journal} {\bibinfo
  {journal} {Phys. Rev. E}\ }\textbf {\bibinfo {volume} {98}},\ \bibinfo
  {pages} {022603} (\bibinfo {year} {2018})}\BibitemShut {NoStop}%
\bibitem [{\citenamefont {Khelfa}\ \emph {et~al.}(2022)\citenamefont {Khelfa},
  \citenamefont {Korbmacher}, \citenamefont {Schadschneider},\ and\
  \citenamefont {Tordeux}}]{Khelfa22}%
  \BibitemOpen
  \bibfield  {author} {\bibinfo {author} {\bibfnamefont {B.}~\bibnamefont
  {Khelfa}}, \bibinfo {author} {\bibfnamefont {R.}~\bibnamefont {Korbmacher}},
  \bibinfo {author} {\bibfnamefont {A.}~\bibnamefont {Schadschneider}},\ and\
  \bibinfo {author} {\bibfnamefont {A.}~\bibnamefont {Tordeux}},\ }\bibfield
  {title} {\bibinfo {title} {Heterogeneity-induced lane and band formation in
  self-driven particle systems},\ }\href
  {https://doi.org/10.1038/s41598-022-08649-4} {\bibfield  {journal} {\bibinfo
  {journal} {Sci. Rep.}\ }\textbf {\bibinfo {volume} {12}},\ \bibinfo {pages}
  {4768} (\bibinfo {year} {2022})}\BibitemShut {NoStop}%
\bibitem [{\citenamefont {Feliciani}\ and\ \citenamefont
  {Nishinari}(2016)}]{Feliciani16}%
  \BibitemOpen
  \bibfield  {author} {\bibinfo {author} {\bibfnamefont {C.}~\bibnamefont
  {Feliciani}}\ and\ \bibinfo {author} {\bibfnamefont {K.}~\bibnamefont
  {Nishinari}},\ }\bibfield  {title} {\bibinfo {title} {Empirical analysis of
  the lane formation process in bidirectional pedestrian flow},\ }\href
  {https://doi.org/10.1103/PhysRevE.94.032304} {\bibfield  {journal} {\bibinfo
  {journal} {Phys. Rev. E}\ }\textbf {\bibinfo {volume} {94}},\ \bibinfo
  {pages} {032304} (\bibinfo {year} {2016})}\BibitemShut {NoStop}%
\bibitem [{\citenamefont {Bacik}\ \emph {et~al.}(2023)\citenamefont {Bacik},
  \citenamefont {Bacik},\ and\ \citenamefont {Rogers}}]{Bacik23}%
  \BibitemOpen
  \bibfield  {author} {\bibinfo {author} {\bibfnamefont {K.~A.}\ \bibnamefont
  {Bacik}}, \bibinfo {author} {\bibfnamefont {B.~S.}\ \bibnamefont {Bacik}},\
  and\ \bibinfo {author} {\bibfnamefont {T.}~\bibnamefont {Rogers}},\
  }\bibfield  {title} {\bibinfo {title} {Lane nucleation in complex active
  flows},\ }\href {https://doi.org/10.1126/science.add8091} {\bibfield
  {journal} {\bibinfo  {journal} {Science}\ }\textbf {\bibinfo {volume}
  {379}},\ \bibinfo {pages} {923} (\bibinfo {year} {2023})}\BibitemShut
  {NoStop}%
\bibitem [{\citenamefont {Vizarim}\ \emph {et~al.}(2025)\citenamefont
  {Vizarim}, \citenamefont {Souza}, \citenamefont {Reichhardt}, \citenamefont
  {Reichhardt}, \citenamefont {Venegas},\ and\ \citenamefont
  {B\'eron}}]{Vizarim25}%
  \BibitemOpen
  \bibfield  {author} {\bibinfo {author} {\bibfnamefont {N.~P.}\ \bibnamefont
  {Vizarim}}, \bibinfo {author} {\bibfnamefont {J.~C.~B.}\ \bibnamefont
  {Souza}}, \bibinfo {author} {\bibfnamefont {C.~J.~O.}\ \bibnamefont
  {Reichhardt}}, \bibinfo {author} {\bibfnamefont {C.}~\bibnamefont
  {Reichhardt}}, \bibinfo {author} {\bibfnamefont {P.~A.}\ \bibnamefont
  {Venegas}},\ and\ \bibinfo {author} {\bibfnamefont {F.}~\bibnamefont
  {B\'eron}},\ }\bibfield  {title} {\bibinfo {title} {Skyrmion-skyrmionium
  phase separation and laning transitions via spin-orbit torque currents},\
  }\href {https://doi.org/10.1103/zq91-42cc} {\bibfield  {journal} {\bibinfo
  {journal} {Phys. Rev. B}\ }\textbf {\bibinfo {volume} {111}},\ \bibinfo
  {pages} {214438} (\bibinfo {year} {2025})}\BibitemShut {NoStop}%
\bibitem [{\citenamefont {Neto}\ and\ \citenamefont {Silva}(2022)}]{Neto22}%
  \BibitemOpen
  \bibfield  {author} {\bibinfo {author} {\bibfnamefont {J.~F.}\ \bibnamefont
  {Neto}}\ and\ \bibinfo {author} {\bibfnamefont {C.~C. d.~S.}\ \bibnamefont
  {Silva}},\ }\bibfield  {title} {\bibinfo {title} {Mesoscale phase separation
  of skyrmion-vortex matter in chiral-magnet--superconductor
  heterostructures},\ }\href {https://doi.org/10.1103/PhysRevLett.128.057001}
  {\bibfield  {journal} {\bibinfo  {journal} {Phys. Rev. Lett.}\ }\textbf
  {\bibinfo {volume} {128}},\ \bibinfo {pages} {057001} (\bibinfo {year}
  {2022})}\BibitemShut {NoStop}%
\bibitem [{\citenamefont {Zarenia}\ \emph {et~al.}(2017)\citenamefont
  {Zarenia}, \citenamefont {Neilson},\ and\ \citenamefont
  {Peeters}}]{Zarenia17}%
  \BibitemOpen
  \bibfield  {author} {\bibinfo {author} {\bibfnamefont {M.}~\bibnamefont
  {Zarenia}}, \bibinfo {author} {\bibfnamefont {D.}~\bibnamefont {Neilson}},\
  and\ \bibinfo {author} {\bibfnamefont {F.~M.}\ \bibnamefont {Peeters}},\
  }\bibfield  {title} {\bibinfo {title} {Inhomogeneous phases in coupled
  electron-hole bilayer graphene sheets: Charge density waves and coupled
  {W}igner crystals},\ }\href {https://doi.org/10.1038/s41598-017-11910-w}
  {\bibfield  {journal} {\bibinfo  {journal} {Sci. Rep.}\ }\textbf {\bibinfo
  {volume} {7}},\ \bibinfo {pages} {11510} (\bibinfo {year}
  {2017})}\BibitemShut {NoStop}%
\bibitem [{\citenamefont {Zhou}\ \emph {et~al.}(2021)\citenamefont {Zhou},
  \citenamefont {Sung}, \citenamefont {Brutschea}, \citenamefont {Esterlis},
  \citenamefont {Wang}, \citenamefont {Scuri}, \citenamefont {Gelly},
  \citenamefont {Heo}, \citenamefont {Taniguchi}, \citenamefont {Watanabe},
  \citenamefont {Zar{\' a}nd}, \citenamefont {Lukin}, \citenamefont {Kim},
  \citenamefont {Demler},\ and\ \citenamefont {Park}}]{Zhou21}%
  \BibitemOpen
  \bibfield  {author} {\bibinfo {author} {\bibfnamefont {Y.}~\bibnamefont
  {Zhou}}, \bibinfo {author} {\bibfnamefont {J.}~\bibnamefont {Sung}}, \bibinfo
  {author} {\bibfnamefont {E.}~\bibnamefont {Brutschea}}, \bibinfo {author}
  {\bibfnamefont {I.}~\bibnamefont {Esterlis}}, \bibinfo {author}
  {\bibfnamefont {Y.}~\bibnamefont {Wang}}, \bibinfo {author} {\bibfnamefont
  {G.}~\bibnamefont {Scuri}}, \bibinfo {author} {\bibfnamefont {R.~J.}\
  \bibnamefont {Gelly}}, \bibinfo {author} {\bibfnamefont {H.}~\bibnamefont
  {Heo}}, \bibinfo {author} {\bibfnamefont {T.}~\bibnamefont {Taniguchi}},
  \bibinfo {author} {\bibfnamefont {K.}~\bibnamefont {Watanabe}}, \bibinfo
  {author} {\bibfnamefont {G.}~\bibnamefont {Zar{\' a}nd}}, \bibinfo {author}
  {\bibfnamefont {M.~D.}\ \bibnamefont {Lukin}}, \bibinfo {author}
  {\bibfnamefont {P.}~\bibnamefont {Kim}}, \bibinfo {author} {\bibfnamefont
  {E.}~\bibnamefont {Demler}},\ and\ \bibinfo {author} {\bibfnamefont
  {H.}~\bibnamefont {Park}},\ }\bibfield  {title} {\bibinfo {title} {Bilayer
  {W}igner crystals in a transition metal dichalcogenide heterostructure},\
  }\href {https://doi.org/10.1038/s41586-021-03560-w} {\bibfield  {journal}
  {\bibinfo  {journal} {Nature (London)}\ }\textbf {\bibinfo {volume} {595}},\
  \bibinfo {pages} {48} (\bibinfo {year} {2021})}\BibitemShut {NoStop}%
\bibitem [{\citenamefont {Vezirov}\ and\ \citenamefont
  {Klapp}(2013)}]{Vezirov13}%
  \BibitemOpen
  \bibfield  {author} {\bibinfo {author} {\bibfnamefont {T.~A.}\ \bibnamefont
  {Vezirov}}\ and\ \bibinfo {author} {\bibfnamefont {S.~H.~L.}\ \bibnamefont
  {Klapp}},\ }\bibfield  {title} {\bibinfo {title} {Nonequilibrium dynamics of
  a confined colloidal bilayer in a planar shear flow},\ }\href
  {https://doi.org/10.1103/PhysRevE.88.052307} {\bibfield  {journal} {\bibinfo
  {journal} {Phys. Rev. E}\ }\textbf {\bibinfo {volume} {88}},\ \bibinfo
  {pages} {052307} (\bibinfo {year} {2013})}\BibitemShut {NoStop}%
\bibitem [{\citenamefont {Lekner}(1991)}]{Lekner91}%
  \BibitemOpen
  \bibfield  {author} {\bibinfo {author} {\bibfnamefont {J.}~\bibnamefont
  {Lekner}},\ }\bibfield  {title} {\bibinfo {title} {Summation of {C}oulomb
  fields in computer-simulated disordered-systems},\ }\href
  {https://doi.org/10.1016/0378-4371(91)90226-3} {\bibfield  {journal}
  {\bibinfo  {journal} {Physica A}\ }\textbf {\bibinfo {volume} {176}},\
  \bibinfo {pages} {485} (\bibinfo {year} {1991})}\BibitemShut {NoStop}%
\bibitem [{\citenamefont {Gr{\o}nbech-Jensen}(1997)}]{GronbechJensen97a}%
  \BibitemOpen
  \bibfield  {author} {\bibinfo {author} {\bibfnamefont {N.}~\bibnamefont
  {Gr{\o}nbech-Jensen}},\ }\bibfield  {title} {\bibinfo {title} {Lekner
  summation of long range interactions in periodic systems},\ }\href
  {https://doi.org/10.1142/S0129183197001144} {\bibfield  {journal} {\bibinfo
  {journal} {Int. J. Mod. Phys. C}\ }\textbf {\bibinfo {volume} {8}},\ \bibinfo
  {pages} {1287} (\bibinfo {year} {1997})}\BibitemShut {NoStop}%
\bibitem [{\citenamefont {Reichhardt}\ and\ \citenamefont
  {Reichhardt}(2022)}]{Reichhardt22}%
  \BibitemOpen
  \bibfield  {author} {\bibinfo {author} {\bibfnamefont {C.}~\bibnamefont
  {Reichhardt}}\ and\ \bibinfo {author} {\bibfnamefont {C.~J.~O.}\ \bibnamefont
  {Reichhardt}},\ }\bibfield  {title} {\bibinfo {title} {Nonlinear dynamics,
  avalanches, and noise for driven {W}igner crystals},\ }\href
  {https://doi.org/10.1103/PhysRevB.106.235417} {\bibfield  {journal} {\bibinfo
   {journal} {Phys. Rev. B}\ }\textbf {\bibinfo {volume} {106}},\ \bibinfo
  {pages} {235417} (\bibinfo {year} {2022})}\BibitemShut {NoStop}%
\bibitem [{\citenamefont {Reichhardt}\ and\ \citenamefont
  {Reichhardt}(2025)}]{Reichhardt25}%
  \BibitemOpen
  \bibfield  {author} {\bibinfo {author} {\bibfnamefont {C.}~\bibnamefont
  {Reichhardt}}\ and\ \bibinfo {author} {\bibfnamefont {C.~J.~O.}\ \bibnamefont
  {Reichhardt}},\ }\bibfield  {title} {\bibinfo {title} {Directional locking
  and hysteresis in stripe- and bubble-forming systems on one-dimensional
  periodic substrates with a rotating drive},\ }\href
  {https://doi.org/10.1103/PhysRevE.111.054119} {\bibfield  {journal} {\bibinfo
   {journal} {Phys. Rev. E}\ }\textbf {\bibinfo {volume} {111}},\ \bibinfo
  {pages} {054119} (\bibinfo {year} {2025})}\BibitemShut {NoStop}%
\bibitem [{\citenamefont {Reichhardt}\ and\ \citenamefont
  {Reichhardt}(2019{\natexlab{c}})}]{Reichhardt19a}%
  \BibitemOpen
  \bibfield  {author} {\bibinfo {author} {\bibfnamefont {C.}~\bibnamefont
  {Reichhardt}}\ and\ \bibinfo {author} {\bibfnamefont {C.~J.~O.}\ \bibnamefont
  {Reichhardt}},\ }\bibfield  {title} {\bibinfo {title} {Active microrheology,
  {H}all effect, and jamming in chiral fluids},\ }\href
  {https://doi.org/10.1103/PhysRevE.100.012604} {\bibfield  {journal} {\bibinfo
   {journal} {Phys. Rev. E}\ }\textbf {\bibinfo {volume} {100}},\ \bibinfo
  {pages} {012604} (\bibinfo {year} {2019}{\natexlab{c}})}\BibitemShut
  {NoStop}%
\bibitem [{\citenamefont {Rothen}\ and\ \citenamefont
  {Piera\ifmmode~\acute{n}\else \'{n}\fi{}ski}(1996)}]{Rothen96}%
  \BibitemOpen
  \bibfield  {author} {\bibinfo {author} {\bibfnamefont {F.}~\bibnamefont
  {Rothen}}\ and\ \bibinfo {author} {\bibfnamefont {P.}~\bibnamefont
  {Piera\ifmmode~\acute{n}\else \'{n}\fi{}ski}},\ }\bibfield  {title} {\bibinfo
  {title} {Mechanical equilibrium of conformal crystals},\ }\href
  {https://doi.org/10.1103/PhysRevE.53.2828} {\bibfield  {journal} {\bibinfo
  {journal} {Phys. Rev. E}\ }\textbf {\bibinfo {volume} {53}},\ \bibinfo
  {pages} {2828} (\bibinfo {year} {1996})}\BibitemShut {NoStop}%
\bibitem [{\citenamefont {Menezes}\ and\ \citenamefont
  {de~Souza~Silva}(2017)}]{Menezes17}%
  \BibitemOpen
  \bibfield  {author} {\bibinfo {author} {\bibfnamefont {R.~M.}\ \bibnamefont
  {Menezes}}\ and\ \bibinfo {author} {\bibfnamefont {C.~C.}\ \bibnamefont
  {de~Souza~Silva}},\ }\bibfield  {title} {\bibinfo {title} {Conformal vortex
  crystals},\ }\href {https://doi.org/10.1038/s41598-017-12807-4} {\bibfield
  {journal} {\bibinfo  {journal} {Sci. Rep.}\ }\textbf {\bibinfo {volume}
  {7}},\ \bibinfo {pages} {12766} (\bibinfo {year} {2017})}\BibitemShut
  {NoStop}%
\bibitem [{\citenamefont {Bellizotti~Souza}\ \emph {et~al.}(2023)\citenamefont
  {Bellizotti~Souza}, \citenamefont {Vizarim}, \citenamefont {Reichhardt},
  \citenamefont {Reichhardt},\ and\ \citenamefont {Venegas}}]{Souza23}%
  \BibitemOpen
  \bibfield  {author} {\bibinfo {author} {\bibfnamefont {J.~C.}\ \bibnamefont
  {Bellizotti~Souza}}, \bibinfo {author} {\bibfnamefont {N.~P.}\ \bibnamefont
  {Vizarim}}, \bibinfo {author} {\bibfnamefont {C.~J.~O.}\ \bibnamefont
  {Reichhardt}}, \bibinfo {author} {\bibfnamefont {C.}~\bibnamefont
  {Reichhardt}},\ and\ \bibinfo {author} {\bibfnamefont {P.~A.}\ \bibnamefont
  {Venegas}},\ }\bibfield  {title} {\bibinfo {title} {Spontaneous skyrmion
  conformal lattice and transverse motion during dc and ac compression},\
  }\href {https://doi.org/10.1088/1367-2630/acd46f} {\bibfield  {journal}
  {\bibinfo  {journal} {New J. Phys.}\ }\textbf {\bibinfo {volume} {25}},\
  \bibinfo {pages} {053020} (\bibinfo {year} {2023})}\BibitemShut {NoStop}%
\bibitem [{\citenamefont {Ben-David}\ \emph {et~al.}(2010)\citenamefont
  {Ben-David}, \citenamefont {Rubinstein},\ and\ \citenamefont
  {Fineberg}}]{BenDavid10}%
  \BibitemOpen
  \bibfield  {author} {\bibinfo {author} {\bibfnamefont {O.}~\bibnamefont
  {Ben-David}}, \bibinfo {author} {\bibfnamefont {S.~M.}\ \bibnamefont
  {Rubinstein}},\ and\ \bibinfo {author} {\bibfnamefont {J.}~\bibnamefont
  {Fineberg}},\ }\bibfield  {title} {\bibinfo {title} {Slip-stick and the
  evolution of frictional strength},\ }\href
  {https://doi.org/10.1038/nature08676} {\bibfield  {journal} {\bibinfo
  {journal} {Nature}\ }\textbf {\bibinfo {volume} {463}},\ \bibinfo {pages}
  {76} (\bibinfo {year} {2010})}\BibitemShut {NoStop}%
\bibitem [{\citenamefont {Tian}\ \emph {et~al.}(2016)\citenamefont {Tian},
  \citenamefont {Tao}, \citenamefont {Yin}, \citenamefont {Zhang},
  \citenamefont {Meng},\ and\ \citenamefont {Tian}}]{Tian16}%
  \BibitemOpen
  \bibfield  {author} {\bibinfo {author} {\bibfnamefont {P.}~\bibnamefont
  {Tian}}, \bibinfo {author} {\bibfnamefont {D.}~\bibnamefont {Tao}}, \bibinfo
  {author} {\bibfnamefont {W.}~\bibnamefont {Yin}}, \bibinfo {author}
  {\bibfnamefont {X.}~\bibnamefont {Zhang}}, \bibinfo {author} {\bibfnamefont
  {Y.}~\bibnamefont {Meng}},\ and\ \bibinfo {author} {\bibfnamefont
  {Y.}~\bibnamefont {Tian}},\ }\bibfield  {title} {\bibinfo {title} {Creep to
  inertia dominated stick-slip behavior in sliding friction modulated by tilted
  non-uniform loading},\ }\href {https://doi.org/10.1038/srep33730} {\bibfield
  {journal} {\bibinfo  {journal} {Sci. Rep.}\ }\textbf {\bibinfo {volume}
  {6}},\ \bibinfo {pages} {33730} (\bibinfo {year} {2016})}\BibitemShut
  {NoStop}%
\end{thebibliography}%

\end{document}